\documentclass[
  reprint,
  doublecolumn,
  aps,
  prd,
  superscriptaddress, 
  nofootinbib,
  amsmath,amssymb
]{revtex4-2}

\usepackage{times, mathrsfs, amsmath, amsfonts, graphics, graphicx, cancel, color, amsthm, bbm, mathtools, amssymb, physics}
\usepackage{upgreek} 
\usepackage{accents}
\usepackage{dcolumn}
\usepackage{bm}
\usepackage{hyperref}
\usepackage{enumerate}
\definecolor{capri}{rgb}{0.0, 0.75, 1.0}

\usepackage{tikz}
\usepackage[normalem]{ulem}
\usepackage{xcolor}
\usetikzlibrary{arrows.meta}

\definecolor{clockorange}{RGB}{190,110,30}
\definecolor{systemblue}{RGB}{10,40,130}

\begin{document}

\preprint{APS/123-QED}

\title{Quantum limits of a space-time reference frame}

\author{Davide Mattei}
\affiliation{Dipartimento di Fisica, Universit\`a di Roma ``Tor Vergata'' and INFN, Sezione di Roma Tor Vergata, Via della Ricerca Scientifica 1, 00133 Roma, Italy}
\affiliation{Institute for Quantum Optics and Quantum Information (IQOQI), Austrian Academy of Sciences, Boltzmanngasse 3, A-1090 Vienna, Austria}

\author{Esteban Castro-Ruiz}
\affiliation{Institute for Quantum Optics and Quantum Information (IQOQI), Austrian Academy of Sciences, Boltzmanngasse 3, A-1090 Vienna, Austria}

\date{\today}

\begin{abstract}

\noindent We study the limitations for defining spatial and temporal intervals when the only available reference frame is a single composite quantum system, whose internal degrees of freedom serve as a temporal reference — a clock — and whose centre-of-mass degrees of freedom act as a spatial reference — a rod. By combining quantum speed limits with the mass–energy equivalence of special relativity, we show that spatial localisability and temporal resolution are not independent: sharpening one inevitably blurs the other. Specifically, the internal-energy coherence needed for precise timekeeping affects the centre-of-mass dynamics, enhancing position spreading during free evolution. As a result, a single composite system cannot act as a perfect quantum reference frame for both space and time, leading to a Heisenberg-like uncertainty relation between spatial and temporal intervals. After analysing this trade-off from an external perspective, we formulate it in a purely relational manner, by means of covariant observables relative to the spatio-temporal quantum reference frame, uncovering an additional intrinsic uncertainty of order the frame’s Compton wavelength.
\end{abstract}

\maketitle

\section{Introduction}

A basic problem in physics is to describe how systems move through space and evolve in time. Operationally, this requires a physical reference frame that locates objects in space and tracks their evolution over time. A natural model for such a reference frame consists of a composite system whose internal degrees of freedom serve as a time reference (a \emph{clock}), while its centre-of-mass (c.o.m.) serves as a spatial reference (a \emph{rod})\footnote{By ``rod'' we mean a reference frame for spatial translations, not for spatial distances.} \cite{clock-latticework-picture}. 
This reference frame naturally splits space-time into space and time: spatial positions are defined relative to the c.o.m., and temporal intervals are determined by the clock’s proper time, consistent with the clock hypothesis \footnote{The clock hypothesis states that the readings of a clock coincide with the proper time predicted by the theory, i.e., the clock couples to the metric field so that its internal evolution faithfully tracks space-time intervals along its worldline.} \cite{clock-hypothesis}. In this way, a single composite system can serve simultaneously as both a spatial and temporal reference.

As far as we know, the physics of clocks and rods is governed by the laws of quantum mechanics (QM), making our composite system a spatio-temporal quantum reference frame (STQRF) \cite{STQRF-Flaminia}.  Investigating the limitations on how well such STQRFs can define spatial and temporal concepts therefore provides a natural route to studying the interplay between quantum theory and space-time physics.

Salecker and Wigner~\cite{Salecker-Wigner} were among the first to take up this challenge, showing that quantum theory imposes restrictions on the mass of a STQRF if it is to define the distance between two space-time events with a given accuracy. This line of reasoning has been extended to account for limitations coming from gravitational phenomena. It has been argued that, if gravitational collapse is to be avoided, there must exist an upper bound on the clock’s mass~\cite{Amelino-Camelia} and on the total number of clock ``ticks'' that can happen within a space-time region \cite{lloyd2012quantum,Quantum-Geometric-Limit}. Relatedly, it has also been proposed that the uncertainty in the metric field induced by the clock itself contributes to the definability of space-time intervals~\cite{metric-fluctuations}, and that gravitational interactions between nearby quantum clocks preclude the joint measurability of time along nearby worldlines~\cite{entanglement-clocks-gravity-Ruiz}.

In this work, we focus on the limitations to the definability of space-time distances with STQRFs that are present already in special relativity (SR). Specifically, combining quantum speed limits with the mass-energy equivalence from SR, we show that the localisability in space and the temporal precision of a STQRF are not independent: improving one inevitably degrades the other.

\begin{figure}[h]
\centering
\includegraphics[width=0.75\linewidth]{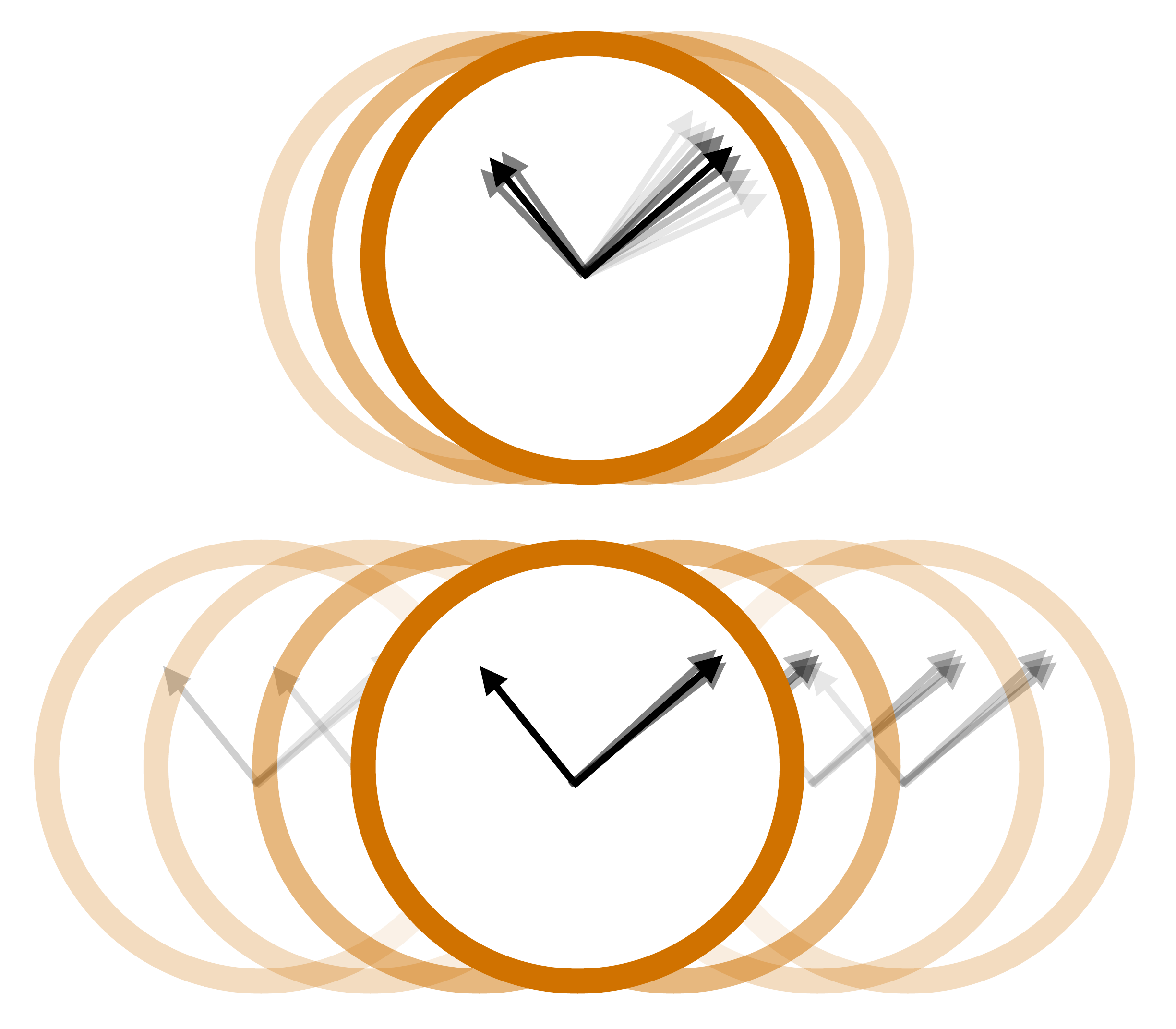}
\caption{Space-time measurability trade-off. Above, a relatively well-localised STQRF remains so at the expense of an increased uncertainty in its clock readings. Below: a sharp clock leads to an increase of position uncertainty upon free evolution, due to the coupling between internal and external degrees of freedom stemming from the clock's dynamical mass (see Eq.~\eqref{eq:evolutionoperator}).} \label{fig:trade-off}
\end{figure}

The effect we study can be explained intuitively as follows (see Fig.~(\ref{fig:trade-off})). The temporal precision of a quantum clock is controlled by the so-called quantum speed limits (QSLs) \cite{MT,MT-bound-Fleming,Margolous-Levitin,Quantum-Limits-Giovannetti,QSL-coherence-Spekkens,QSL-coherence-geom,QSL-entanglement}. In particular, the Mandelstam–Tamm bound ~\cite{MT,Margolous-Levitin}, states that the orthogonalization time $t_\perp$ — the minimal time for a state to evolve into an orthogonal one — is inversely proportional to the internal–energy spread:

\begin{equation}\label{MT-bound-intro}
t_\perp \;\ge\; \frac{\pi \hbar}{2\,\Delta_\Psi \hat{H}_{c}},
\end{equation}
where $\Delta_\Psi^2 \hat{H}_{c}=\langle \hat{H}_{c}^2\rangle_\Psi-\langle \hat{H}_{c}\rangle_\Psi^2$ is the variance of the clock Hamiltonian in the state $\ket{\Psi}$. By the mass–energy equivalence, the internal energy contributes to the total mass. Assuming this holds at the quantum level, the mass is promoted to an operator \cite{QEEP-Zych,decoherence-time-dilation}
\begin{equation}\label{eq:massoperator}
\hat{m} \;=\; m \;+\; \frac{\hat{H}_{c}}{c^2}.
\end{equation}
Because the mass governs the c.o.m. dynamics, the same energy spread that increases time precision inevitably affects position uncertainty during its evolution.

After introducing our clock model in Sec.~\ref{sec:model}, in Sec. \ref{sec:external-perspective} we analyse the interplay between spatial localization and time precision from an \emph{external} perspective. We revisit the Salecker–Wigner thought experiment \cite{Salecker-Wigner} by considering the clock as a \emph{composite} particle, showing that internal-energy uncertainty enhances uncertainty in the c.o.m. position. Combining this effect with the Mandelstam–Tamm bound yields an operational trade-off: spatial localisability and temporal precision cannot be simultaneously optimized. Thus, a material clock cannot serve as an infinitely precise reference for both space and time, a result that can be expressed by a Heisenberg-like uncertainty relation. In Sec.\ref{sec:internal-perspective} we go beyond the external perspective to study how physics ``looks like'' from the perspective of a QRF \cite{Giacomini2019QRF, Vanrietvelde-change-of-perspective, QRF-trinity, Castro-relative-subsystems, carette2025operational}. We develop a \emph{relational} framework, where the position and time evolution a quantum system under study are described relative to the STQRF, without relying on an external space-time reference frame. The c.o.m. is used to define a covariant position observable \cite{QRF-QI,QRF-Loveridge} of the system relative to the STQRF, while the clock degrees of freedom provide the temporal parameter with respect to which covariant observables evolve. It is in this context that the interplay between spatial and temporal localisability becomes fundamental, since both quantities now determine the uncertainty in the relative position between the system and the STQRF. What we find is that, compared to the usual external perspective, the relational formulation yields an additional contribution to the (relative) position uncertainty of the order of the Compton wavelength of the STQRF.

\section{Model and notation}\label{sec:model}

\noindent In this work, we consider a \emph{composite} quantum system with Hilbert space $\mathcal{H}_{\mathrm{rc}} = \mathcal{H}_\mathrm{r} \otimes \mathcal{H}_\mathrm{c}$, 
where $\mathcal{H}_\mathrm{r} $ describes the c.o.m. degrees of freedom, 
serving as a spatial reference (the rod), 
and $\mathcal{H}_\mathrm{c}$ describes the internal degrees of freedom, which provide a temporal reference frame (the clock). We will refer to it as a quantum reference frame for space and time, or \emph{STQRF}.

We work in a regime where quantum-field–theoretic effects such as pair creation are negligible. In this regime, the leading relativistic corrections can be incorporated by extending the mass–energy equivalence principle to the quantum level, that is, by promoting the rest mass to an operator that includes the internal energy contribution~\cite{visibility-proper-time,decoherence-time-dilation,Zych-thesis}. As a result, the free Hamiltonian couples the internal and kinematical degrees of freedom:
\begin{equation}\label{eq:H_SR_exact}
    \hat{H}_{\mathrm{rc}} \;=\; \frac{\hat{p}^2}{2\hat{m}} \;+\; \hat{m} c^2,
\end{equation}
where \(\hat{m}\) is given by Eq.~(\eqref{eq:massoperator}). Note that \(\hat{m}\) acts only on \(\mathcal{H}_c\), while \(\hat{x}\) and \(\hat{p}\) act only on \(\mathcal{H}_r\).
Consequently, \([\hat{m},\hat{x}] = [\hat{m},\hat{p}] = 0\), and expressions such as \(\hat{p}\hat{m}^{-1}\) are well defined.

This model was originally proposed to study quantum time dilation effects in weak gravitational fields~\cite{visibility-proper-time}
and has been later derived from various approaches (see e.g. ~\cite{decoherence-time-dilation}). A complete review can be found in~\cite{Zych-thesis}.

The special-relativistic time dilation effect becomes manifest once the mass operator is expanded in powers of \(\hat{H}_\mathrm{c}/(m c^2)\),
assuming that the internal energy scale is much smaller than the rest energy \(m c^2\) \footnote{
Formally, the low-energy regime requires that the spectrum of the internal Hamiltonian \(\hat{H}_\mathrm{c}\) lies well below \(mc^2\), or that we restrict to a subspace of the Hilbert space where this condition is satisfied.
This is a natural assumption when considering special-relativistic corrections to non-relativistic QM.
}.
To first order,
\begin{equation}\label{eq:H_SR_expanded}
    \hat{H}_{rc} \;=\; \frac{\hat{p}^2}{2m} \;+\; m c^2 \;+\; \hat{H}_\mathrm{c}\!\left(1 - \frac{\hat{p}^2}{2 m^2 c^2}\right)\!.
\end{equation}
This expression shows that the clock’s evolution depends on the kinematical motion via the SR time dilation (or redshift) factor \footnote{
At first order in \(c^{-2}\), the SR factor is \(\gamma^{-1} = \sqrt{1 - v^2/c^2} \simeq 1 - v^2/(2c^2)\).
} \cite{Smith-Ahmadi-relativistic-clocks, Khandelwal-Lock-Woods}.
To emphasise that the clock runs according to its proper time, which depends on its momentum, it is useful to write explicitly the evolution operator
\begin{equation}\label{eq:evolutionoperator}
    \hat{U}_{rc}(t)
    \;=\; e^{-\frac{i}{\hbar}t\,\frac{\hat{p}^2}{2m}}\;
          e^{-\frac{i}{\hbar}\,\hat{\tau}(\hat{p})\,\hat{H}_\mathrm{c}},
\end{equation}
where \({\tau}(\hat{p}) \coloneq t\,\Delta(\hat{p})\) is the (operator-valued) proper time and \(\Delta(\hat{p}) = 1 - \hat{p}^2/2m^2c^2\) is the (first-order) time dilation factor.

In the non-relativistic limit \(c \!\to\! \infty\),
the internal and external degrees of freedom evolve independently, with the clock thus tracking ``absolute'' Newtonian time. 

In the following, we focus on how the STQRF evolves in time. Our analysis goes beyond the effect studied by Salecker and Wigner \cite{Salecker-Wigner} by including the entanglement between the clock and its c.o.m. induced by the mass operator of Eq.~\eqref{eq:H_SR_exact}.

\section{Interplay between position- and time-uncertainty}\label{sec:external-perspective}

\noindent We consider a STQRF described by the Hamiltonian~\eqref{eq:H_SR_exact}, which evolves freely and is used to define spatial and temporal intervals. We study the limitations that QM imposes on how sharply these quantities can be defined. In our analysis, we use the perspective of an external observer who can perform measurements on the STQRF. In section \ref{sec:internal-perspective}, we adopt a relational perspective, studying how the STQRF defines the temporal and spatial coordinates of other systems, and the limitations thereof.

\subsection{Spatial uncertainty}

We begin by recalling the seminal work by Salecker and Wigner. They considered a STQRF which, during its free-evolution, is used to measure space-like distances. A light pulse is sent from the STQRF to a mirror and reflected back so that, measuring the elapsed time $t$ with the clock, gives the spatial separation, $L = c t / 2$, between its c.o.m. and the mirror \footnote{This setup assumes an approximately flat region, i.e., the distance $L$ involved is small compared with the curvature of space.}. They then considered the limitations imposed by QM on the definability of such a space-like interval. In particular, due to Heisenberg uncertainty, the STQRF cannot be arbitrarily localized in space \footnote{Salecker and Wigner also analysed the recoil
(``kickback'') of the light signal on the clock, which in general
introduces an additional contribution to the distance uncertainty.
In the present work we focus on the limitation arising from the quantum spreading of the clock’s
wave packet.}. A particle with initial position spread $\sigma$ has a minimal velocity spread $\delta v \gtrsim \hbar/(2m\sigma)$, leading to $\delta x^2 \gtrsim \sigma^2 + \left({\hbar\,t}/{2m\sigma}\right)^2$.
Minimizing over $\sigma$ yields the well-known bound, also known as ``Standard Quantum Limit" (SQL), on the measurement of space-like distances:
\begin{equation}\label{eq:Salecker-Wigner-bound}
    \delta x \gtrsim \sqrt{\frac{\hbar |t|}{m}} .
\end{equation}

We now analyse aspects of this thought experiment within the model presented in Section \ref{sec:model}. Specifically, we study the uncertainty in the c.o.m. position after it evolves in time by means of the Hamiltonian of Eq. ~\eqref{eq:H_SR_exact}. In addition to the limitation analysed in \cite{Salecker-Wigner}, we describe new limitations, arising from the impact of the clock's internal-energy spread.

In the Heisenberg picture, the position operator evolves under the Hamiltonian ~\eqref{eq:H_SR_exact} as
\begin{equation}\label{eq:X-operator-Heisenberg-picture}
    \hat{x}(t) \;=\; \hat{x} \;+\; \frac{\hat{p}}{\hat{m}}\,t, 
\end{equation}
where $t$ is the time coordinate corresponding to a space-time foliation in the external perspective. For a quantum state \(\ket{\Psi}_{rc}\in\mathcal{H}_r\otimes\mathcal{H}_c\),
we denote the variance of an operator \(\hat{A}\) by \(\Delta_\Psi^2 \hat{A}=\langle \hat{A}^2\rangle_\Psi-\langle \hat{A}\rangle_\Psi^2\)
and the covariance of operators $\hat{A}$ and $\hat{B}$ by \(\mathrm{Cov}_\Psi(\hat{A},\hat{B})=\tfrac{1}{2}\langle \hat{A}\hat{B}+\hat{B}\hat{A}\rangle_\Psi-\langle \hat{A}\rangle_\Psi\langle \hat{B}\rangle_\Psi\).
From Eq.~\eqref{eq:X-operator-Heisenberg-picture} we directly obtain
\begin{equation}\label{eq:X-variance-free}
    \Delta_\Psi^2 \hat{x}(t) \;=\; \Delta_\Psi^2 \hat{x} \;+\; 2t\,\mathrm{Cov}_\Psi(\hat{x},\hat{v}) \;+\; t^2\,\Delta_\Psi^2 \hat{v} \, ,
\end{equation}
where the velocity operator is \(\hat{v}=\tfrac{1}{i\hbar}[\hat{x},\hat{H}] = \hat{p}\,\hat{m}^{-1}\). Note that the mass enters this definition as an operator on $\mathcal{H}_c$.

A non-trivial lower bound at any time \(t\) in Eq.~\eqref{eq:X-variance-free} can only be obtained by assuming \emph{non-contractive} preparations, i.e.\ \(\mathrm{Cov}_\Psi(\hat{x},\hat{v})\ge0\).
This assumption, implicit also in the derivation of the Salecker–Wigner bound~\eqref{eq:Salecker-Wigner-bound}, is natural in the operational scenario we consider. It is well known that correlations in phase space can transiently reduce space–time uncertainty. In principle, this can be used to beat the SQL on two consecutive measurements of the position of a massive particle \cite{SQL-Contractive-states}. However, to do so likely requires an external measuring apparatus to control and prepare the state after the first measurement \footnote{Moreover, at present it remains an experimental challenge, since it is still not known if and how the current theoretical proposals can be practically implemented \cite{Quantum-Geometric-Limit}.}. This work, instead, focuses on the intrinsic limitations of a composite quantum system used as the \emph{only} reference for both space and time (see Section \ref{sec:internal-perspective}). It is therefore natural to restrict attention to symmetric states satisfying \(\mathrm{Cov}_\Psi(\hat{x},\hat{v})=0\). In Appendix \ref{app:contractive-states}, we discuss the nature of non-symmetric states, their relation with the SQL and to our work.

Under this condition, we can find a general lower bound to the position spread using the Heisenberg uncertainty relation
\begin{equation}\label{eq:Heis-Unc-Rel-XV}
   \Delta_\Psi^2 \hat{x}\,\Delta_\Psi^2 \hat{v}
   \;\ge\; \frac{\hbar^2}{4}\,\big\langle \hat{m}^{-1}\big\rangle_\Psi^2 ,
\end{equation}
which follows from \([\hat{x},\hat{v}]=i\hbar\,\hat{m}^{-1}\). Combining Eqs.~\eqref{eq:X-variance-free} and \eqref{eq:Heis-Unc-Rel-XV} gives
\begin{equation}\label{eq:variance-X-not-minimized}
  \Delta_\Psi^2 \hat{x}(t)
  \;\ge\;
  \left(\frac{\hbar}{2}\,\big\langle \hat{m}^{-1}\big\rangle_\Psi\right)^{\!2}
  \frac{1}{\Delta_\Psi^2 \hat{v}}
  \;+\; t^2\,\Delta_\Psi^2 \hat{v} \, .
\end{equation}
Minimizing with respect to \(\Delta_\Psi^2 \hat{v}\) yields
\begin{equation}\label{eq:variance-X-free-minimized}
   \Delta_\Psi^2 \hat{x}(t) \;\ge\; \hbar |t| \,\big\langle \hat{m}^{-1}\big\rangle_\Psi \, .
\end{equation}
This inequality generalizes the Salecker–Wigner bound \eqref{eq:Salecker-Wigner-bound} by including the dynamical effects of the mass, encoded in the mass operator \(\hat{m}\), acting on the STQRF's internal degrees of freedom.

The role of internal-energy spread becomes manifest once we expand \(\hat{m}^{-1}\) in powers of \(\hat{H}_\mathrm{c}/(mc^2)\). Up to second order, Eq.~\eqref{eq:variance-X-free-minimized} becomes

\begin{equation}\label{eq:X-variance-free-avg-energy}
   \Delta_\Psi^2 \hat{x}(t)
   \,\geq\,
   \frac{\hbar|t|}{m}
   \left(
   1-\frac{\langle \hat{H}_\mathrm{c}\rangle_\Psi}{mc^2}
   +\frac{\Delta_\Psi^2 \hat{H}_\mathrm{c}+\langle \hat{H}_\mathrm{c}\rangle_\Psi^2}{m^2c^4}
   \right).
\end{equation}
Equation~\eqref{eq:X-variance-free-avg-energy} shows that internal-energy spread \(\Delta_\Psi^2 \hat{H}_\mathrm{c}\) enhances the spreading of the wavepacket,
while the mean internal energy \(\langle \hat{H}_\mathrm{c}\rangle_\Psi\) contributes with the opposite sign.
This is natural, since \(\langle \hat{H}_\mathrm{c}\rangle_\Psi / c^2\) effectively renormalizes the rest mass and thereby reduces the spreading. Re-defining the rest mass as \(\overline{m} \coloneq m+\langle \hat{H}_\mathrm{c}\rangle_\Psi / c^2\),
we can isolate the effect of the energy spread, and Eq.~\eqref{eq:X-variance-free-avg-energy} becomes
\begin{equation}\label{eq:X-variance-free-final}
   \Delta_\Psi^2 \hat{x}(t)
   \,\geq\,
   \frac{\hbar|t|}{\overline{m}}
   \;+\; \frac{\hbar|t|}{m}
   \frac{\Delta_\Psi^2 \hat{H}_\mathrm{c}}{m^2c^4}.
\end{equation}
This bound is saturated only by a specific class of states — those minimizing the \(\hat{x}_r\)–\(\hat{v}_r\) uncertainty relation~\eqref{eq:Heis-Unc-Rel-XV} — first analysed in~\cite{Zych-MUS-states} and referred to as ``minimum-uncertainty-states" (MUS). These are entangled states of the form

\begin{equation}\label{eq:MUS-C-state}
   \ket{\Psi} \;=\; \sum_i \phi_i \int dp\; \psi_{v_0,x_0}(p,\epsilon_i)\,\ket{p}_{r}\otimes\ket{\epsilon_i}_{c}.
\end{equation}
The Gaussian wavepacket $\psi_{v_0,x_0}(p,\epsilon_i)$, written explicitly in Appendix \ref{app:MUS-states}, is correlated with the internal energy of the clock through the relativistic mass $m_i=m+\epsilon_i/c^2$, such that, for a given internal-energy spread, the resulting spatial delocalization is minimal. In Appendix~\ref{app:MUS-states}, we compare MUS states with standard Gaussian states -- those that minimize the $\hat{x}_r$--$\hat{p}_r$ uncertainty relation. We show explicitly
how, for Gaussian states, the uncertainty in the internal-energy leads to a greater spatial spreading than the MUS of Eq. \eqref{eq:MUS-C-state}, in agreement with the results obtained in \cite{Zych-MUS-states}.

In summary, the spatial uncertainty of the STQRF is controlled by its internal dynamics:  
(i) it grows with the spread of internal energy, and  
(ii) it is such that achieving minimal uncertainty at fixed energy spread necessarily entails entanglement between the clock and its c.o.m. In the following, we show that these two ingredients—energy uncertainty and entanglement—also set the fundamental limit on the STQRF's temporal uncertainty.

\subsection{Temporal uncertainty}
For a clock prepared in a pure state and evolving unitarily, time precision can be unambiguously defined by the orthogonalization time \(t_\perp\):  
the minimal time required for the state to evolve into a perfectly distinguishable one. QM imposes fundamental bounds on this quantity, known as the quantum speed limits (QSLs). In particular, we focus on the Mandelstam–Tamm \cite{MT,Margolous-Levitin} bound, roughly stating that $t_\perp$ is inversely proportional to the energy spread of the clock’s Hamiltonian. More precisely,
\begin{equation}\label{eq:MT-bound}
    t_\perp \;\ge\; \frac{\pi\hbar}{2\,\Delta_\Psi \hat{H}_\mathrm{c}}.
\end{equation}

For mixed states, in general, there is no unique notion of distinguishability, and therefore no natural analogue of $t_\perp$. Accordingly, several extensions of the QSL have been proposed by defining a suitable distance in the space of density operators—such as the \emph{trace distance}~\cite{QSL-coherence-Spekkens,QSL-coherence-geom} or the \emph{Bures distance} via fidelity~\cite{Quantum-Limits-Giovannetti,QSL-entanglement,QSL-Taddei-QFI}—and asking for the minimal time, $t_\delta(\hat{\rho}_c)$, required for the state \(\hat{\rho}_c(t)\) to move by a fixed distance \(\delta\) from \(\hat{\rho}_c(0)\) \footnote{ For a recent review of the topic, we refer to \cite{QSL-review-DeffnerCampbell}}. As an example, in Ref.~\cite{QSL-coherence-Spekkens}, assuming unitary evolution, the authors find a bound in terms of the trace distance:
\begin{equation}\label{eq:MT-mixed}
    t_\delta(\hat{\rho}_c) \,\geq\, \frac{\hbar\,\delta}{F_{\hat{H}_c}(\hat{\rho}_c)}\,, 
\end{equation}
where

\begin{gather}\label{eq:MT-mixed-F}
    F_{\hat{H}_c}(\hat{\rho}_c) \coloneq \|\big[\hat{\rho}_c,\hat{H}_c\big]\|_1 
\end{gather}
and $\|\hat{O} \|_1 = \mathrm{tr}\!\Big[\sqrt{\hat{O}^\dagger \hat{O}}\,\,\Big]$ is the trace norm. The quantity defined in \eqref{eq:MT-mixed-F} quantifies the coherence of \(\hat{\rho}_c\) in the eigenbasis of the Hamiltonian \(\hat{H}_c\). This is evident by expanding the commutator, $[\hat{\rho}_c,\hat{H}_c] \;=\; \sum_{ij} (\epsilon_j - \epsilon_i)\, \hat{\rho}_c^{\,ij} \,\ketbra{i}{j},$ which leads (via the triangle inequality) to

\begin{gather}\label{eq:MT-mixed-decoherence}
     F_{\hat{H}_c}(\hat{\rho}_c) \;\leq\; \sum_{i,j} |\epsilon_j - \epsilon_i|\,|\hat{\rho}_c^{\,ij}|.
\end{gather}
Hence, as the off-diagonal elements in the energy basis, \(\hat{\rho}_c^{ij}\),  decay, \(t_\delta(\hat{\rho}_c)\) increases (hence clock's precision decreases).

Physically, any amount of mixing in the clock’s state degrades its temporal resolution, which fundamentally relies on quantum coherence. Indeed, QSLs for mixed states yield tighter bounds that reduce to the Mandelstam–Tamm form when the state is pure. For instance, one can show that
$
F_{\hat{H}_c}(\hat{\rho}_c)\;\le\;2\,\Delta_{\hat{\rho}_c} \hat{H}_c,
$
where \(\Delta_{\hat{\rho}_c} \hat{H}_c\) is the energy spread with respect to \(\hat{\rho}_c\) ~\cite{QSL-coherence-Spekkens}. Equality holds only for pure states, recovering Eq.~\eqref{eq:MT-bound}.

In our framework, the clock corresponds to the internal degrees of freedom of a composite particle $\hat{\rho}_c = \mathrm{tr}_r[\hat{\rho}_{rc}]$. Decoherence in the internal-energy basis arises from entanglement between the clock and its c.o.m. (the rod) through the relativistic contribution of internal energy to the mass operator \(\hat{m}=m+\hat{H}_\mathrm{c}/c^2\).  
Even if the clock and the rod are initially uncorrelated, they become entangled under the Hamiltonian~\eqref{eq:H_SR_exact}; consequently, the internal sector decoheres during free evolution.

In Appendix~\ref{app:clock-precision}, we compute the off-diagonal elements of the clock’s reduced density matrix for a generic discrete clock obtaining, up to order $o\,\Big(\big(\Delta \epsilon_{ij}/mc^2\big)^{\!2}\Big)$
\begin{equation}\label{eq:clock-off-diag-terms}
|\hat{\rho}_c^{\,ij}(t)|
= |\hat{\rho}_c^{\,ij}(0)|
\Bigg[1-\frac{1}{2}
\Bigg(\frac{t}{2\hbar m }\frac{\Delta \epsilon_{ij}}{mc^2}\Bigg)^{\!2}
\Delta^2_r(\hat{p}^2_{ij})\Bigg],
\end{equation}
where \(\Delta\epsilon_{ij}\) is the difference between energy levels $i$ and $j$, $\hat{\rho}_c^{ij}(0) = \displaystyle\int_{\mathbb{R}} dp\; \bra{p} \hat{\rho}_{rc}^{ij} \ket{p}_r$ and $\hat{\rho}_{rc}^{ij} = \bra{\epsilon_i}\hat{\rho}_{rc}\ket{\epsilon_j}_c$. The quantity
$
\Delta_r\!\big(\hat{p}^2_{ij}\big)
:= \big\langle \hat{p}_{ij}^4\big\rangle_r
   - \big(\big\langle \hat{p}^2_{ij}\big\rangle_r\big)^2
$
is the variance of $\hat p^2$ with respect to the momentum ``cross
distribution'' associated with the pair $(i,j)$: for any function $f$ we define the normalized expectation value:
\begin{gather}\label{eq:SQL-cross-avg}
    \langle \big(f(\hat{p})\big)_{ij}\rangle_r:=\frac{\int_{\mathbb{R}} dp\,\bra{p}\hat{\rho}_{rc}^{ij}\ket{p}_r\, f(p)}{\int_{\mathbb{R}} dp\,\bra{p}\hat{\rho}_{rc}^{ij}\ket{p}_r},
\end{gather}
Equation~\eqref{eq:clock-off-diag-terms} reveals two distinct sources of decoherence:
(i)~initial entanglement, through \(|\hat{\rho}_c^{\,ij}(0)|\leq1\); and  (ii)~``dynamical'' entanglement, induced by the quantum uncertainty in the SR time dilation factor  
\(\Delta(\hat{p})=1-\hat{p}^2/2m^2c^2\) under time evolution, which yields the time-dependent term.

A QSL can be found by generalizing Eq.~\eqref{eq:MT-mixed-F} \footnote{More general extensions of QSL to non-unitary dynamics have been studied in the literature, e.g. in \cite{QSL-delCampo-open},\cite{QSL-Deffner-Lutz}.} to the present case, where the clock does not evolve under a single unitary, but undergoes a random-unitary dephasing, that is,
\begin{gather}\label{eq:clock-mixture-unitaries}
    \hat{\rho}_c(t)
    = \int_{\mathbb{R}} dp\, e^{-\frac{i}{\hbar} t\,\Delta(p) \hat{H}_\mathrm{c}}\,
    \bra{p}\hat{\rho}_{rc}\ket{p}_r\,
    e^{\frac{i}{\hbar} t\,\Delta(p) \hat{H}_\mathrm{c}}.
\end{gather}
In Appendix~\ref{app:clock-precision} we show that, in this case, Eq.~\eqref{eq:MT-mixed-F} can be replaced by
\begin{equation}\label{eq:MT-mixed-definition-general}
     F_{\hat{H}_\mathrm{c}}(\hat{\rho}_c)
     = \int_{\mathbb{R}} dp\;\Delta(p)\;
     \big\|\,[\bra{p}\hat{\rho}_{rc}\ket{p}_r,\hat{H}_\mathrm{c}]\,\big\|_1.
\end{equation}
Similarly to Eq.~\eqref{eq:MT-mixed-decoherence}, in this case we have

\begin{gather}
  F_{\hat{H}_\mathrm{c}}(\hat{\rho}_c)
  \leq  \sum_{ij} |\epsilon_i - \epsilon_j| \;
  \Big|\int_{\mathbb{R}} dp\,\Delta(p)\, \bra{p}\hat{\rho}_{rc}^{\,ij}\ket{p}_r\Big|
  \nonumber\\[2pt]
  = \sum_{ij} |\epsilon_i - \epsilon_j|\;|\hat{\rho}_c^{\,ij}(0)|
  \left(1- \frac{\big\langle\hat{p}^2_{ij}\;\big\rangle_r}{2m^2c^2}\right).
  \label{eq:QSL-generalized-static}
\end{gather}

Eq. \eqref{eq:QSL-generalized-static} shows that the temporal precision is reduced by (i) the loss of coherence due to the initial entanglement 
$|\hat{\rho}_c^{\,ij}(0)|$, (ii) special-relativistic time dilation. Intuitively, temporal precision depends on the rate-of-evolution of the clock (see App. \ref{app:clock-precision}) which is slowed down by the time dilation factor $\tau \to t = \tau \Delta(\hat{p})$. 
As an example, in the case where the clock is a qubit, $i=0,1$, we have $|\epsilon_i - \epsilon_j| = 2\Delta_{\Psi}\hat{H}_c$ and the QSL reads: 

\begin{gather}
    t_\delta \geq \frac{\delta \hbar}{2\Delta_\Psi \hat{H}_c} \frac{1}{|\hat{\rho}_c^{01}(0)|}  \left(1- \frac{\big\langle\hat{p}^2_{01}\;\big\rangle_r}{2m^2c^2}\right)
    \label{eq:QSL-qubit-static}
\end{gather}

In Appendix~\ref{app:clock-precision} we also derive a tighter bound that takes into account the dynamical loss of coherence, i.e.\ the time-dependent part of Eq.~\eqref{eq:clock-off-diag-terms}. This refinement does not change the qualitative conclusions of this section, namely that time precision is fundamentally determined by quantum coherence in the internal-energy basis:
(i) the fundamental limit for pure states is given by the Mandelstam--Tamm bound, which shows that the precision in time is inversely proportional to the internal-energy spread; and (ii) any loss of coherence in the internal-energy basis further reduces temporal precision.

\subsection{Trade-off between space and time}

From the above analysis, we observe that there exists a trade-off between spatial and temporal precision: sharpening one inevitably blurs the other. In the following, we make this statement precise.

The most direct ``space-time uncertainty relation'' follows from combining the position spread~\eqref{eq:X-variance-free-final} with the Mandelstam–Tamm bound~\eqref{eq:MT-bound}, which is valid also in the mixed-state case, since any degree of mixing can only \emph{increase} the minimal distinguishability time \footnote{
For notational simplicity, we denote the clock’s precision time by \(t_\perp\) even if the clock’s state is not pure—then to be understood as a distance-threshold time rather than a strict orthogonalization time.}.
Multiplying Eq.~\eqref{eq:X-variance-free-final} by \(t_\perp^2\) yields
\begin{equation}\label{eq:trade-off-free}
   \Delta_\Psi \hat{x}(t)\,t_\perp
   \;\ge\;
   \frac{\pi}{2}\,
   \sqrt{\frac{\hbar |t|}{\overline{m}}}\,
   \frac{\hbar}{m c^2},
\end{equation}
where $\overline{m} = m+\langle \hat{H}_\mathrm{c}\rangle_\Psi / c^2$. Equation~\eqref{eq:trade-off-free} expresses the core result of this work: a composite particle cannot serve as a perfectly sharp reference for both space and time. The smaller \(t_\perp\) is, corresponding to higher temporal precision, the larger the spatial uncertainty \(\Delta_\Psi \hat{x}(t)\). Thus, improving time resolution unavoidably degrades spatial localization—and vice versa.

Another aspect of the space--time interplay concerns entanglement in the initial state of the STQRF. Achieving
minimum spatial uncertainty at a given internal-energy spread
$\Delta_{\Psi}\hat{H}_c$ (Eq.~\eqref{eq:X-variance-free-final}) requires
entanglement between the clock and its c.o.m.; however, our analysis of temporal
uncertainty shows that the same entanglement induces decoherence in the
internal-energy basis and thus worsens temporal precision (Eqs. \eqref{eq:QSL-generalized-static} and \eqref{eq:QSL-qubit-static}). As a concrete
illustration, in Appendix~\ref{app:trade-off} we compare the MUS defined in
Eq.~\eqref{eq:MUS-C-state} with standard Gaussian states for a qubit clock and we summarize this behaviour in terms
of quantitative trade-offs between spatial uncertainties and QSLs: the advantage that MUS have over Gaussian states when it comes to spatial precision is compensated by the advantage Gaussian states have over MUS when it comes to temporal precision. This suggests that, once QSLs are taken into account, the trade-off in Eq.~\eqref{eq:trade-off-free} should be viewed as a weak form of a more general, tighter space--time trade-off.

In summary, the results derived here show that spatial localization and time precision for a composite system are not independent. On the one hand, the internal-energy spread that enhances the ability of the clock to define time intervals simultaneously delocalizes its c.o.m. in space. On the other hand, the quantum correlations that enhance localisability in space for a given internal-energy spread, simultaneously limit the definability of time. Neither aspect—spatial localization nor temporal resolution—can be made arbitrarily precise without compromising the other.

\section{Relational description}\label{sec:internal-perspective}

\noindent In the previous section, we considered a composite system evolving in time with respect to an external, abstract reference frame. As time passes, the c.o.m. wave packet spreads in position space. This spreading is further enhanced by the spread in internal energy (see Eq.~\eqref{eq:X-variance-free-final}), 
which is inversely proportional to the clock’s precision in timekeeping. 
This implies a trade-off between how sharply a composite particle is localized in space and how precise it can be as a timekeeper (Eq.~\eqref{eq:trade-off-free}).

This fact becomes more significant when the composite system is effectively used as a quantum reference frame \cite{QRF-QI,QRF-Loveridge,Giacomini2019QRF,QRF-trinity}, for both space and time. In particular, we observe that the external perspective implicitly assumes an ideal clock defining the classical time parameter $t$. Thus, \textit{a priori}, there's no reason to require that the clock be a precise temporal reference -- and thus have a large energy spread. It is only when the clock is effectively used as a QRF to define time evolution that the trade-off (Eq.\eqref{eq:trade-off-free}) becomes essential. 

Building on this observation, in this section we introduce a simple toy model in which the STQRF is used to describe the temporal and spatial parameters governing the quantum state of another physical system. We then show how the interplay between space and time appears in the uncertainty of the relational position operator—that is, the position operator in the Heisenberg picture defined with respect to the STQRF rather than an abstract external reference frame. In doing so, we uncover an additional source of uncertainty in position arising from the clock degrees of freedom.

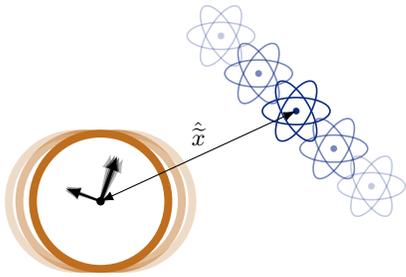
\begin{figure}[t]
\centering
\begin{tikzpicture}[>=Latex, line cap=round, line join=round, scale=1]

\foreach \dx/\opa/\aone/\atwo in {
   0.18/0.60/80/150,  
  -0.18/0.60/60/170,  
   0.32/0.25/75/145,  
  -0.32/0.25/65/175}  
{
  \begin{scope}[shift={(\dx,0)}, opacity=\opa]
    \fill[white] (0,0) circle (0.9);
    \draw[line width=3pt, clockorange] (0,0) circle (0.9);
    \draw[very thick, black] (0,0) -- (\aone:0.65);
    \draw[thick,      black] (0,0) -- (\atwo:0.50);
    \fill (0,0) circle (1.6pt);
  \end{scope}
}

\fill[white] (0,0) circle (0.9);
\draw[line width=3pt, clockorange] (0,0) circle (0.9);

\foreach \ang/\opa in {
  67/0.55,  
  73/0.55,  
  62/0.30,  
  78/0.30   
}{
  \draw[very thick, black, -{Latex[length=2mm]}, opacity=\opa]
    (0,0) -- (\ang:0.65);
}

\foreach \ang/\opa in {
  157/0.40,
  163/0.40
}{
  \draw[thick, black, -{Latex[length=2mm]}, opacity=\opa]
    (0,0) -- (\ang:0.50);
}

\draw[very thick, black, -{Latex[length=2mm]}] (0,0) -- (70:0.65);   
\draw[thick,      black, -{Latex[length=2mm]}] (0,0) -- (160:0.50);  
\fill (0,0) circle (1.6pt);


\foreach \dx/\dy/\opa in {
      0/0/1,
      0.5/-0.5/0.55,
      1.0/-1.0/0.25,
     -0.5/0.5/0.55,
     -1.0/1.0/0.25}{
  \begin{scope}[shift={(2.6+\dx,1.2+\dy)}, opacity=\opa]

    \fill[systemblue] (0,0) circle (0.045);

    \draw[systemblue, line width=0.6pt] (0,0) ellipse (0.45 and 0.18);
    \draw[systemblue, line width=0.6pt, rotate=60]  (0,0) ellipse (0.45 and 0.18);
    \draw[systemblue, line width=0.6pt, rotate=120] (0,0) ellipse (0.45 and 0.18);

  \end{scope}
}


\draw[black, <->]
  (0,0) -- (2.6,1.2)
  node[midway, above, yshift=1pt, black] {$  \hat{\widetilde{x}}$};

\end{tikzpicture}
\caption{Relational description of a quantum system (the atom) with respect to
the STQRF. The spatial and temporal references are provided,
respectively, by the c.o.m.\ and the internal degrees of freedom, both subject
to quantum uncertainty principles. As a consequence, their quantum uncertainty
impacts the relational position observable $ \hat{\widetilde{x}}$ between the STQRF
and the system, limiting our ability to jointly locate the system sharply in space and
time.}
\label{fig:stqrf-clock}
\end{figure}

\subsection{Model -- ideal clock}
We model a STQRF as a composite system with Hilbert space $\mathcal{H}_{rc}=\mathcal{H}_r\otimes\mathcal{H}_c$ as a quantum reference frame for the one-dimensional groups of spatial and time translations. The c.o.m. degrees of freedom $\mathcal{H}_r$ serve as a spatial reference (the rod), while the internal degrees of freedom $\mathcal{H}_c$ serve as a temporal reference (the clock).

We consider an ``ideal'' clock, whose Hilbert space is $\mathcal{H}_c = \mathcal{L}^2(\mathbb{R})$ and whose Hamiltonian $\hat{H}_c$ has an unbounded spectrum $\mathrm{spec}(\hat{H}_c) = \mathbb{R}$ \cite{QRF-trinity,Giovannetti-quantum-time,quantizing-time-Smith,Castro-quantum-clocks-temporal-localisability}.
Although unphysical, this model has often been used in the literature because it captures the essential features of a quantum-clock while simplifying computations. 
An ideal clock is characterized by a self-adjoint (proper-) time operator $\hat{\tau}_c$ that is canonically conjugate to the internal Hamiltonian,
$ [\hat{\tau}_c, \hat{H}_c] = i\hbar.$
The canonical commutation relation leads directly to a time-energy uncertainty relation,
\begin{gather}\label{MT-bound-continuous-clock}
    \Delta_\Psi\hat{\tau}_c\,\Delta_\Psi \hat{H}_c \geq \frac{\hbar}{2},
\end{gather}
which plays the role of a QSL (Eq.~\eqref{eq:MT-bound}).

As in the previous section, the c.o.m. degree of freedom is that of a single particle, with Hilbert space $\mathcal{H}_r = \mathcal{L}^2(\mathbb{R})$. Thus, the most general pure state of the STQRF, $\ket{\Psi}_{rc}\in\mathcal{H}_{rc}$, has the form:
\begin{gather}\label{eq:QRF-state-general}
     \ket{\Psi}_{rc} =  \int_{\mathbb{R}^2} d\epsilon\,dp \;
     \Psi(p,\epsilon)\,\ket{p}_r \otimes \ket{\epsilon}_c.
\end{gather}
 Physically, any entanglement between the clock and its c.o.m.\ arises from the fact that the internal energy contributes to the total mass $m(\epsilon) = m_r + \epsilon/c^2$, which in turn can enter the c.o.m. state. Therefore, it's natural to consider wavepackets in the form
\begin{gather}\label{eq:QRF-state-wavepacket}
    \Psi(p,\epsilon) = \psi_{v_0}(p,\epsilon)\,\phi_{\epsilon_0}(\epsilon),
\end{gather}
where $\psi_{v_0}(p,\epsilon)$ is the wave packet in momentum space that potentially depends on $\epsilon$ through the mass $m(\epsilon)$ (see App. \ref{app:MUS-states} for an explicit example), and $\phi_{\epsilon_0}(\epsilon)$ is the wave packet in energy space with an associated uncertainty $\Delta_\Psi \hat{H}_c$, which is the parameter of interest. For instance, consider the Gaussian wavepacket
\begin{gather}\label{eq:Gaussian-clock-state}
    \phi_{\epsilon_0}(\epsilon)
    =  \left(\frac{1}{2\pi\,\Delta_\Psi^2 \hat{H}_c}\right)^{\!1/4}
    \exp\!\left[-\frac{(\epsilon-\epsilon_0)^2}{4\,\Delta_\Psi^2 \hat{H}_c}\right].
\end{gather}

Note that, even though the clock is ideal, in the sense that it carries the regular representation of the group of time-translations  \cite{Castro-relative-subsystems,QRF-Anne-Catherine,Loveridge-relative-quantum-time, Hoehn-Vanrietvelde-swithing-clocks} it is not infinitely precise, as states have in general a finite energy uncertainty. An infinitely precise clock would correspond to the limit $\Delta_\Psi \hat{H}_c \to \infty$.

\subsection{Relational position operator}

We now use the STQRF to define the position of another free quantum particle with Hilbert space $\mathcal{H}_s = \mathcal{L}^2(\mathbb{R})$. We assume that the system and the STQRF are uncorrelated and do not interact.\footnote{It is well known that entanglement between the QRF and the system can decrease the uncertainty of relative observables~\cite{Quantum-Geometric-Limit}. However, in this work we are interested in the use of the STQRF in generic situations, and thus it is natural to require that it be uncorrelated from the system.}

To distinguish the QRF’s c.o.m.\ (the rod) from the system, we use the subscripts $r$ and $s$, respectively. 
The generators of space and time translations on the QRF are, respectively, the total momentum $\hat{p}_{\mathrm{r}}$ and the Hamiltonian $\hat{H}_{rc}$ of Eq.~\eqref{eq:H_SR_exact}. Similarly, the generators of space and time translations for the system are its momentum $\hat{p}_s$ and Hamiltonian $\hat{H}_s = \hat{p}_s^{\,2}/(2m_s)$.
For $(x,t)\in\mathbb{R}^2$, the group of space–time translations acts as
\begin{equation}\label{eq:QRF-joint-group-action}
\hat{U}_{\mathrm{rc}}(x,t) \otimes \hat{U}_s(x,t)
= e^{-\tfrac{i}{\hbar}\big(x\,\hat{p}_{\mathrm{r}}+t\,\hat{H}_{\mathrm{rc}}\big)}
\otimes 
e^{-\tfrac{i}{\hbar}\big(x\,\hat{p}_s+t\,\hat{H}_s\big)}.
\end{equation}

Starting from the system position operator $\hat{x}_s$, we construct the corresponding relational observable:
\begin{gather}\label{eq:relational-position-operator-general}
  \hat{\widetilde{x}}
  = \int_{\mathbb{R}^2} dx\,dt\; \hat{E}(x,t)\,\otimes\,\hat{U}_s(x,t)\,  \hat{x}_s\, \hat{U}^\dagger_s(x,t).
\end{gather}
Here $\hat{E}(x,t)$ are the elements of a POVM on the STQRF that is covariant under space–time translations~\eqref{eq:QRF-joint-group-action}, i.e.\ it satisfies
\begin{equation}
\hat{U}_{\mathrm{rc}}(x',t')\,\hat{E}(x,t)\,\hat{U}_{\mathrm{rc}}^\dagger(x',t')
= \hat{E}(x+x',t+t') \, .
\end{equation}
Physically, the coordinates $(x,t)$ labelling the system’s observables become \emph{relational}: they are specified with respect to the STQRF via the POVM $\hat{E}(x,t)$. As a result, the operator~\eqref{eq:relational-position-operator-general} is invariant under global space–time translations defined in Eq. \eqref{eq:QRF-joint-group-action}.

A covariant POVM can be constructed from a “seed’’ element $\hat{E}_0>0$ as
\begin{equation}\label{eq:POVM-from-seed}
\hat{E}(x,t) = \hat{U}_{\mathrm{rc}}(x,t)\,\hat{E}_0\,\hat{U}^\dagger_{\mathrm{rc}}(x,t).
\end{equation}
Substituting Eq. \eqref{eq:POVM-from-seed} into
Eq.~\eqref{eq:relational-position-operator-general} shows that $\hat{\widetilde{x}}$ is obtained by a
G-twirl (or incoherent group average) over the joint space–time translations \cite{QRF-QI}.  
Here, we choose
\begin{equation}\label{eq:POVM-seed-all-orders}
\hat{E}_{0}\;=\;\hat{\Delta}^{1/2}\,\Big(\ketbra{x_0}{x_0}_{r}\otimes\ketbra{\tau_0}{\tau_0}_{c}\Big)\,\hat{\Delta}^{1/2},
\end{equation}
where $\hat{\Delta}$ is the time–dilation (redshift) operator, whose matrix elements $\Delta(p, \epsilon)$ are defined via the spectral function
$E(p,\epsilon)=\bra{p}_r\bra{\epsilon}_c\,\hat{H}_{\mathrm{rc}}\,\ket{p}_r\ket{\epsilon}_c
=\frac{p^2}{2\,m(\epsilon)}+m(\epsilon)c^2$,
as
\begin{gather}
\Delta(p,\epsilon)=\frac{\partial E(p,\epsilon)}{\partial \epsilon}
=\;1-\frac{p^{2}}{2\,m(\epsilon)^{2} c^{2}},
\label{definition-time-dilation-factor}
\end{gather}
with $m(\epsilon) = m_r + \epsilon/c^2$. We restrict to the subspace where $m(\epsilon)c\gg |p|$, so that $\hat{\Delta}$ is positive. This restriction is natural, since we are considering SR corrections to non-relativistic quantum mechanics\footnote{
For Gaussian states, which have unbounded support, this condition is only approximately satisfied. Formally, one should project states onto the $\hat{\Delta} > 0$ subspace.}.
Since $\hat{\Delta}$ depends on both internal and external degrees of freedom, it does not commute with the sharp projectors $\ketbra{x_0}{x_0}_{r}$ and $\ketbra{\tau_0}{\tau_0}_{c}$. The symmetric “sandwich’’ in Eq.~\eqref{eq:POVM-seed-all-orders} resolves ordering ambiguities. In Appendix~\ref{app:covariant-POVM} we show that, with this choice of seed, the resulting POVM $\hat{E}(x,t)$ is correctly normalised, i.e.
\(\int_{\mathbb{R}^2} dx\,dt\,\hat{E}(x,t) = \mathbb{I}_r\otimes\mathbb{I}_c\).

From this POVM, in Appendix~\ref{app:relative-position-operator} we compute the relational position operator defined in Eq.~\eqref{eq:relational-position-operator-general}, which reads
\begin{align}
     \hat{\widetilde{x}}_{(x_0,\tau_0)}
=& \, \hat{x}_s
+ \frac{\hat{p}_s}{m_s}\,\Bigl\{\hat{\tau}_c-\tau_0,\;\hat{\Delta}^{-1}\Bigr\}\nonumber \\
\;-&\;
\Bigl(
\hat{x}_r
+ \hat{p}_{\mathrm{r}}\,
\Bigl\{\hat{\tau}_c-\tau_0,\;(\hat{m}\,\hat{\Delta})^{-1}\Bigr\}
- x_0
\Bigr),
\label{eq:relational-position-rel}
\end{align}
where $\{\hat{A},\hat{B}\}=\tfrac{1}{2}(\hat{A}\hat{B}+\hat{B}\hat{A})$. In the non-relativistic limit $c\to\infty$ one has $\hat{\Delta}\to\mathbb{I}_c \otimes \mathbb{I}_r$ and $\hat{m}\to m_r\,\mathbb{I}_c$. Therefore, the “seed’’ element \eqref{eq:POVM-seed-all-orders} reduces to a sharp projector in position and time space, and the relational position operator reduces to
\begin{gather}
  \hat{\widetilde{x}}^{\,\,\text{n.r.}}_{(x_0,\tau_0)} 
= \hat{x}_s + \frac{\hat{p}_s}{m_s}\,(\hat{\tau}_c-\tau_0)
\;-\;
\Bigl(\hat{x}_r + \frac{\hat{p}_{\mathrm{r}}}{m_r}\,(\hat{\tau}_c-\tau_0) - x_0\Bigr),
\label{eq:relational-position-non-rel}
\end{gather}
which is the relative position between the system and the STQRF in the Heisenberg picture, with the rod and the clock conditioned on the readout $x_0$ and $\tau_0$, respectively. Comparing \eqref{eq:relational-position-non-rel} with \eqref{eq:relational-position-rel}, we see that the proper time of the clock is corrected by the time dilation factor ($\hat{\tau}_c\to \hat{t}=\hat{\tau}_c\,\hat{\Delta}^{-1}$), and the mass of the rod by the internal energy ($m_r\to \hat{m}$). The anti-commutators resolve ordering ambiguities for terms involving both $\hat{\tau}_c$ and $\hat{m}$.

\subsection{Relational-position uncertainty}

Given the relational position operator~\eqref{eq:relational-position-rel}, we can now compute the associated uncertainty. In particular, our goal is to determine a minimal bound for this quantity, similarly to what was done for the (Heisenberg) position operator of the STQRF with respect to an external reference frame in Eq.~\eqref{eq:X-variance-free-final}. The key difference is that the external time parameter is now replaced by the time according to the clock, which has an intrinsic uncertainty \(\Delta^2_\Psi \hat{\tau}_c\) (see Eq.~\eqref{MT-bound-continuous-clock}). 

The variance of the operator~\eqref{eq:relational-position-rel} is defined as
\begin{gather}\label{eq:def-variance-Dirac-position}
    \Delta_\Psi^2 \,  \hat{\widetilde{x}}_{(x_0,\tau_0)}
    \;=\; \big\langle  \hat{\widetilde{x}}_{(x_0,\tau_0)}^{\,2} \big\rangle_\Psi 
      \;-\; \big\langle  \hat{\widetilde{x}}_{(x_0,\tau_0)} \big\rangle_\Psi^{2},
\end{gather}
where \(\ket{\Psi} \in \mathcal{H}_{rc} \otimes \mathcal{H}_s\) is the joint state of the STQRF and the system. To gain physical insight, it is useful to consider first the non-relativistic limit~\eqref{eq:relational-position-non-rel}. In Appendix~\ref{app:relational-position-spread} we show that \footnote{As in the rest of the paper, we assume all states to be symmetric in phase space, so that \(\mathrm{Cov}(\hat{x},\hat{p})=0\). We discuss this assumption in Appendix~\ref{app:contractive-states}.}
\begin{align}
\Delta_\Psi^2 \,  \hat{\widetilde{x}}^{\,\text{n.r.}}_{(x_0,\tau_0)}
      \;=&\; \Delta_\Psi^2 \hat{x}_s \;+\; \Delta_\Psi^2 \hat{x}_r \nonumber \\
      \hspace{5mm}
         +& \Biggl(\frac{\Delta_\Psi^2 \hat{p}_s}{m_s^2}
                 + \frac{\Delta_\Psi^2 \hat{p}_r}{m_r^2}\Biggr)
           \Bigl(\tau_0^2 + \Delta_\Psi^2 \hat{\tau}_c\Bigr)
\nonumber \\
        \hspace{5mm}
        +& \Biggl(\frac{\langle \hat{p}_s\rangle_\Psi}{m_s}
                -\frac{\langle \hat{p}_r\rangle_\Psi}{m_r}\Biggr)^{\!2}
           \Delta_\Psi^2 \hat{\tau}_c .
\label{eq:spread-relational-position-non-rel-full}
\end{align}
We observe that the spread of the relational position operator contains the sum of the position variances of both the system $s$ and the rod $r$, consistent with the fact that, already in the non-relativistic limit, considering covariant observables  typically leads to greater uncertainties than in standard quantum theory \cite{Jorquera-Riera-Loveridge-PhaseSpaceQRF}. Importantly, the external time parameter is replaced by the clock’s proper time \(\tau_0\) and its associated uncertainty \(\Delta_\Psi^2 \hat{\tau}_c\), which also determines the uncertainty in the average relative motion (last term of Eq.~\eqref{eq:spread-relational-position-non-rel-full}). 

To isolate the contribution arising solely from the STQRF, we neglect the uncertainty due to the system (equivalently, we take the limit \(m_s \rightarrow \infty\)) and also ignore the uncertainty due to the average relative motion (last line of \eqref{eq:spread-relational-position-non-rel-full}). These two sources of uncertainty only increase the total uncertainty. This leads to the inequality
\begin{gather}
\Delta_\Psi^2 \,  \hat{\widetilde{x}}^{\,\text{n.r.}}_{(x_0,\tau_0)}
\;\ge\; \Delta_\Psi^2 \hat{x}_r
\;+\; \frac{\Delta_\Psi^2 \hat{p}_r}{m_r^2}
\Bigl(\tau_0^2 + \Delta_\Psi^2 \hat{\tau}_c\Bigr).
\label{eq:spread-relational-position-non-rel-rod}
\end{gather}
Using the commutation relation between $\hat{x}_r$ and $\hat{p}_r$, the minimum uncertainty associated with the relational position operator is
\begin{gather}
\Delta_\Psi^2 \,  \hat{\widetilde{x}}^{\,\text{n.r.}}_{(x_0,\tau_0)}
\;\ge\; \frac{\hbar}{m_r}\,\sqrt{\tau_0^2 + \Delta_\Psi^2 \hat{\tau}_c}.
\label{eq:minimum-spread-relational-position-non-rel-rod}
\end{gather}
This result is directly analogous to the Salecker–Wigner bound~\eqref{eq:Salecker-Wigner-bound} for a free particle, with the external time parameter replaced by the clock’s proper time and its intrinsic uncertainty. Importantly, even if no time evolution has occurred according to the clock ($\tau_0=0$), the spread is still bounded from below by the clock’s temporal uncertainty, $\Delta_\Psi \hat{\widetilde{x}}^{\,\text{n.r.}}_{(x_0,0)} \ge (\hbar/m_r)\,\sqrt{\Delta^2_\Psi \hat{\tau}_c}$. In the non-relativistic regime this minimal spread can, in principle, vanish in the idealised limit of a perfect clock, $\Delta^2_\Psi \hat{\tau}_c \to 0$, which formally requires an unbounded internal-energy spread, $\Delta^2_\Psi \hat{H}_c \to \infty$. We expect this to no longer hold once relativistic effects are included, since the internal-energy spread then influences the dynamics of the clock, as discussed in Sec.~\ref{sec:external-perspective}. Our goal is therefore to derive an inequality for the full operator~\eqref{eq:relational-position-rel}.

In Appendix~\ref{app:relational-position-spread} we consider corrections up to order \(o(c^{-4})\), and we find the following inequality:

\begin{gather}
     \Delta_\Psi^2 \,  \hat{\widetilde{x}}_{(x_0,\tau_0)} \;\gtrsim\; 
     \Delta_\Psi^2 \hat{x}_r \;+\; \Delta_\Psi^2 \hat{v}_r\,\tau_0^2 
     \;+\; \frac{2}{3}\,\frac{\Delta_\Psi^2 \hat{p}_{\mathrm{r}}}{m_r^2}\,\Delta_\Psi^2 \hat{\tau}_c,
 \label{eq:spread-relational-position-rel-rod}
\end{gather}
where \(\hat{v}_r = \hat{p}_{\mathrm{r}}/\hat{m}\) and $\hat{m}^{-1}$ is understood as its series truncated at \(O(c^{-4})\). The symbol \(\gtrsim\) indicates that neglected higher-order terms are positive or subleading in this expansion \footnote{In particular, correlations between the clock observables lead to negative contributions that are, however, subleading in our Taylor expansion. We refer to App.\ref{app:relational-position-spread} for the details.}. The only assumption made on the state \(\Psi\) is \(\mathrm{Cov}(\hat{p}_{\mathrm{r}}^{\,2},\hat{\tau}_c^{\,2})\ge0\), which physically means that the spreads in momentum and in clock time are positively correlated. This is justified because the clock’s internal Hamiltonian couples to its c.o.m. momentum: uncertainty in momentum induces decoherence in the internal-energy basis, thereby reducing the clock’s precision (see Appendix~\ref{app:clock-precision}).

It now becomes evident that the time–energy uncertainty relation (Eq. \eqref{MT-bound-continuous-clock}) plays a crucial role 
in the uncertainty of the relational position. As in the non-relativistic case, there is a term proportional to \(\Delta_\Psi^2 \hat{\tau}_c\), accounting for the uncertainty in the clock’s proper time. In particular, when \(\tau_0 \simeq 0\) and the dynamics is negligible, one has:

\begin{gather}\label{eq:spread-relational-position-init-times}
  \Delta_\Psi^2 \,  \hat{\widetilde{x}}_{(x_0,0)} \;\gtrsim\; 
  \sqrt{\frac{2}{3}}\,\frac{\hbar}{m_r}\,\Delta_\Psi \hat{\tau}_c.
\end{gather}
where $\Delta_\Psi \hat{\tau}_c \coloneq \sqrt{\Delta^2_\Psi \hat{\tau}_c}$. However, taking the limit \(\Delta^2_\Psi \hat{\tau}_c \to 0\) (i.e. \(\Delta^2_\Psi \hat{H}_c \to \infty\)) is not harmless now, since the internal-energy spread affects the dynamics at later times. Indeed, from Eq. \eqref{eq:spread-relational-position-rel-rod} we have


\begin{gather}
    \Delta_\Psi^2 \,  \hat{\widetilde{x}}_{(x_0,\tau_0)}
    \;\gtrsim\; \Delta_\Psi^2 \hat{x}_r \;+\; \Delta_\Psi^2 \hat{v}_r\,\tau_0^2,
\end{gather}
Using the commutation relations between \(\hat{x}_r\), \(\hat{v}_r\) (see Eq. \eqref{eq:Heis-Unc-Rel-XV}) and expanding 
\(\big\langle \hat{m}^{-1}\big\rangle_\Psi \simeq \tfrac{1}{\overline{m}_r}\bigl(1+\tfrac{\Delta_\Psi^2 \hat{H}_\mathrm{c}}{m_r^2 c^4}\bigr)\) we get:

\begin{equation}\label{eq:variance-Dirac-position-energy spread}
   \Delta_\Psi^2 \,  \hat{\widetilde{x}}_{(x_0,\tau_0)}
   \;\gtrsim\; \frac{\hbar |\tau_0|}{\overline{m}_r} \;+\; 
   \frac{\hbar |\tau_0|}{m_r}\,
   \frac{\Delta_\Psi^2 \hat{H}_{c}}{m_r^2 c^4},
\end{equation}

This inequality is completely analogous to Eq.~\eqref{eq:X-variance-free-final}, computed in the previous section using an external reference frame. The difference is that here \(|\tau_0|\) has an operational meaning -- it is the time according to the clock -- whereas \(t\) in Eq.~\eqref{eq:X-variance-free-final} refers to an external time parameter. Combining Eq. \eqref{eq:variance-Dirac-position-energy spread} with the time-energy uncertainty relation (Eq. \eqref{MT-bound-continuous-clock}), gives

\begin{equation}\label{eq:trade-off-relational-position}
   \Delta_\Psi \,  \hat{\widetilde{x}}_{(x_0,\tau_0)}\; \Delta_\Psi \hat{\tau}_c
   \;\gtrsim\; \frac{1}{2}\sqrt{\frac{\hbar |\tau_0|}{\overline{m}_r}}\,
   \frac{\hbar}{m_r c^2}.
\end{equation}
This result is analogous to the trade-off~\eqref{eq:trade-off-free}, showing that the STQRF cannot serve as a perfectly sharp reference for both space and time. But now it is derived \emph{relationally}, without assuming access to an external reference frame for space or time.

Therefore, both the uncertainty in proper time $\Delta_\Psi^2 \hat{\tau}_c$ and the uncertainty in the internal energy $\Delta_\Psi^2 \hat{H}_c$ contribute to the position spread in Eq.~\eqref{eq:spread-relational-position-rel-rod}. Explicitly, using the commutation relations between \(\hat{x}_r\), \(\hat{v}_r\), and \(\hat{p}_{\mathrm{r}}\), and expanding 
\(\big\langle \hat{m}^{-1}\big\rangle_\Psi \simeq \overline{m}_r^{-1}\bigl(1+\tfrac{\Delta_\Psi^2 \hat{H}_\mathrm{c}}{m_r^2 c^4}\bigr)\) we find:

\begin{gather}
     \Delta_\Psi^2 \,  \hat{\widetilde{x}}_{(x_0,\tau_0)} \;\gtrsim\; 
     \Delta_\Psi^2 \hat{x}_r 
     + \left(\frac{\hbar}{2\,\overline{m}_r}\,\tau_0\right)^{\!2}
       \left(1+\frac{\Delta_\Psi^2 \hat{H}_\mathrm{c}}{m_r^2 c^4}\right)^{\!2}
       \frac{1}{\Delta_\Psi^2 \hat{x}_r} \nonumber \\[3pt]
     \hspace{12mm}
     + \frac{2}{3}\left(\frac{\hbar}{2 m_r}\right)^{\!2}\,\Delta^2_\Psi \hat{\tau}_c
       \frac{1}{\Delta_\Psi^2 \hat{x}_r}.
\label{eq:variance-Dirac-bound}
\end{gather}
Since time and energy spreads are inversely related (see Eq. \ref{MT-bound-continuous-clock}), the minimal uncertainty in the relational position can be found by minimizing with respect to one of them. At sufficiently large times \(|\tau_0| \gg \hbar/(m_r c^2)\) — when the internal-energy spread begins to play a role in the dynamics (see App. \ref{app:relational-position-spread}) — the minimum leads to:

\begin{gather}\label{eq:rel-pos-spread-final}
        \Delta_\Psi^2 \,  \hat{\widetilde{x}}_{(x_0,\tau_0)} \;\gtrsim\;
        \frac{\hbar}{\overline{m}_r}\,|\tau_0| 
        +\frac{1}{\sqrt{3}} \left( \frac{\hbar}{m_r c}\right)^2.
\end{gather}
The first term corresponds to the usual quantum spreading of a free particle of mass \(\overline{m}_r\), identified here with the rod, while the second term arises from the internal degrees of freedom of the QRF — the clock. Therefore, the quantum nature of the clock introduces an additional contribution to the relational position uncertainty, of the order of the Compton wavelength of the rod.

\section{Discussion}

Taking seriously the idea that space and time are what rods and clocks measure we have studied the limitations to the definability of space-time intervals by means of quantum reference frames for space and time. We found that, already in the regime where only leading relativistic effects are relevant, a single composite system cannot serve as an ideal reference for both space and time. The reason is simple: increasing the clock’s precision requires coherence (superpositions) between internal-energy eigenstates; however, through mass–energy equivalence, this energy spread affects the centre-of-mass dynamics, leading to greater spatial delocalisation under free evolution. Furthermore, for a given internal-energy spread, maximal localisability in space requires entanglement between the internal and c.o.m., which in turn reduces clock's precision in time, as constrained by quantum speed limits. Temporal precision and spatial localisability are therefore \emph{not independent}: sharpening one inevitably blurs the other. 

We explored the consequences of this trade-off within the framework of quantum reference frames (QRFs) and relational observables. In this setting, a composite system serves as a QRF for both space and time -- a STQRF -- describing the dynamics of another system in a fully relational manner: the spatial coordinate is replaced by a quantum position observable, and the external time parameter by a quantum time observable. Here, the interplay between spatial and temporal localisability becomes fundamental, since one simultaneously requires a precise clock and a well-localized centre of mass—a condition that, as demonstrated in the first part, cannot be satisfied. In particular, we computed the relational position operator and found that the quantum nature of the clock introduces an additional contribution to its uncertainty, of the order of the Compton wavelength of the particle.

A natural next step is to include gravity in the model. While we expect that placing the STQRF in an external, weak gravitational field (as in \cite{visibility-proper-time}) would not lead to substantial modifications, we do not expect this to remain true once the STQRF itself acts as a gravitating source. We therefore leave for future work the extension to gravitational interactions between the clock and the system, or between multiple clocks, and the study of how the effects uncovered here are modified in such scenarios, building on recent works~\cite{entanglement-clocks-gravity-Ruiz,Castro-quantum-clocks-temporal-localisability}.

It is of fundamental importance to test regimes where both quantum and relativistic features of clocks become relevant. In recent years, several proposals have suggested that such regimes may be within experimental reach (see e.g. \cite{zych-trapped-atoms-interf}, \cite{pikowski-optical-ion-clocks}). These experiments will ultimately clarify the range of validity of the model adopted here and, consequently, of our results.

On the conceptual side, this work suggests that treating clocks as real physical systems, rather than as idealised objects, may lead to new insights into the phenomena that arise at the interface between quantum mechanics and space-time.

\begin{acknowledgments}
We thank Luca Apadula for insightful discussions.
\end{acknowledgments}

\bibliography{apssamp}

\appendix

\renewcommand{\thesubsection}{\thesection.\arabic{subsection}}
\renewcommand{\thesubsubsection}{\thesubsection.\alph{subsubsection}}

\section{Standard quantum limit and contractive states}\label{app:contractive-states}

In this appendix, we discuss in detail the relation between contractive states, the SQL and the bounds presented in the main text. 

The Salecker–Wigner bound is a representative instance of a standard quantum limit (SQL): it follows when one neglects position–momentum correlations that, in principle, can transiently reduce the position uncertainty \cite{SQL-Contractive-states}, \cite{Quantum-Geometric-Limit}. This is explicit in our derivation \eqref{eq:X-variance-free}, where we assumed $\mathrm{Cov}_\Psi(\hat{x}, \hat{v})\!\ge 0$, thereby excluding contractive preparations. In particular, from \eqref{eq:X-variance-free} we see that if $\mathrm{Cov}_\Psi(\hat{x}, \hat{v})<0$ one can have a transient contraction, meaning a finite time-interval in which the position uncertainty decreases as the quantum state evolves. Specifically, for

\begin{gather}\label{eq_app:contractive-window}
    |t|\;\lesssim\;\tau_c:=\frac{|\mathrm{Cov}_\Psi(\hat{x}, \hat{v})|}{\Delta_\Psi^2 \hat{v}}.
\end{gather}
Beyond $|t|>\tau_c$, the quadratic (spreading) term dominates and $\Delta_\Psi^2 \hat{x}(t)$ increases monotonically.
While the trade-off presented in Eq. \eqref{eq:trade-off-free} relies on wave packets in their expanding regime, allowing for contractive states with $\mathrm{Cov}_\Psi(\hat{x}, \hat{v}) < 0$ makes the inequality of Eq. \eqref{eq:variance-X-free-minimized} essentially trivial, leading to $\Delta_\Psi \hat{x}(t) \geq 0$. Furthermore, we observe that the energy spread magnifies the contraction. The reason is that a spread in the internal energy ``amplifies the native trend'': whatever the packet would do without an internal-energy spread (expand or contract), the spread will make it do so \emph{even more}. To see this, take a separable initial state $\ket{\Psi}=\ket{\psi}_{ r}\otimes\ket{\phi}_{c}$, so that
\begin{equation}
\mathrm{Cov}_\Psi(\hat{x}, \hat{v})=\langle \hat{m}^{-1}\rangle_\Psi\,\mathrm{Cov}_{\Psi}(\hat{x},\hat{p}).
\end{equation}
From \eqref{eq:X-variance-free}, at early times (before the Gaussian spreading dominates):

\begin{equation}\label{eq:app-linear-trend}
\Delta_\Psi^2 \hat{x}(t)\;\simeq \Delta_\Psi^2 \hat{x}(0) + \;2t\,\langle \hat{m}^{-1}\rangle_\Psi\,\mathrm{Cov}_{\Psi}(\hat{x},\hat{p}),
\end{equation}
and, choosing $\mathrm{Cov}_{\Psi}(\hat{x},\hat{p})\leq0$ to enforce contraction, the expansion of $\hat{m}^{-1}$ yields

\begin{equation}
\Delta_\Psi^2 \hat{x}(t)\;\simeq \Delta_\Psi^2\hat{x}(0) -\frac{2t}{\overline{m}}\, \left| \mathrm{Cov}_{\Psi}(\hat{x},\hat{p})\right|  \left( 1+  \,\frac{\Delta_\Psi^2 \hat{H}_c}{m^2c^4} \right),
\end{equation}

showing that the term depending on the energy variance exacerbates the contraction. 

The discovery of contractive states~\cite{SQL-Contractive-states} initiated a debate. Caves observed that a fundamental bound emerges when one considers two consecutive position measurements ~\cite{SQL-Defence-Caves}. This follows from the Heisenberg-picture commutator for a free particle between $\hat{x}(t)$ and $\hat{x}(0)$, namely $\left[\hat{x}(t),\hat{x}(0)\right] = i \hbar t/m$, which implies:
\begin{gather}
    \Delta_\Psi \hat{x}(t) \; \Delta_\Psi \hat{x}(0) \geq \frac{\hbar \,|t| }{2m}.
\end{gather}
Thus the two uncertainties $\Delta_\Psi \hat{x}(t)$ and $\Delta_\Psi \hat{x}(0)$ cannot be simultaneously smaller than $\sqrt{\hbar |t|/2m}$. Caves concluded that the SQL remains unavoidable if one demands that both the initial and final spreads be limited. Intuitively, correlations between $\hat{x}$ and $\hat{p}$ require a large initial spread $\Delta_\Psi \hat{x}(0)$ (in phase space: a squeezed and rotated distribution, e.g. a squeezed ellipse along the $-\pi/4$ direction).
This argument can be easily extended to our model. Using Eq. \eqref{eq:X-operator-Heisenberg-picture} we find $\left[\hat{x}(t),\hat{x}(0)\right] = i\hbar t/\hat{m}$, which implies

\begin{gather}
    \Delta_\Psi \hat{x}(t) \; \Delta_\Psi \hat{x}(0) \geq \frac{\hbar\, |t| }{2}\langle \hat{m}^{-1}\rangle_\Psi
\end{gather}
An expansion of $\hat{m}^{-1}$ up to order $(\hat{H}_c / mc^2)^2$ leads to 

\begin{gather}
    \Delta_\Psi \hat{x}(t) \; \Delta_\Psi \hat{x}(0) \gtrsim \frac{\hbar \,|t| }{2\overline{m}}\left( 1+  \,\frac{\Delta_\Psi^2 \hat{H}_c}{m^2c^4} \right).
\end{gather}
This inequality is the analogue, for the product of the spreads
at two different instants of time, of
Eq.~\eqref{eq:trade-off-free}, and it shows explicitly that the internal-energy uncertainty enhances this product of spreads.

Later on, Ozawa pointed out that, in a sequence of two position measurements, the term $\Delta_\Psi\hat x(0)$, corresponds to the uncertainty in position \emph{after} the first measurement. Importantly, this does not
necessarily coincide with the uncertainty in the measurement readout
itself. Ozawa proposed a measurement scheme in which (i) the initial state of the system and probe is such that the uncertainty in the first position measurement  can in principle be smaller than $\sqrt{\hbar |t|/m}$, and
(ii) the interaction with the probe is engineered so that the
post-measurement state of the particle is contractive. If the second
position measurement is performed within the contractive window (Eq. \eqref{eq_app:contractive-window}), the
SQL can then be beaten.

The ongoing debate (see, e.g.,~\cite{Quantum-Geometric-Limit}) concerns the
physical realizability of such protocols, which typically require
active feedback, i.e.\ preparing the system in a state conditioned on
the measurement outcome. Under such conditions, it is in principle possible to
achieve precisions beyond the SQL in the measurement of spacelike distances.

The main reason why we assume non-contractive states in our work is that we
look at a \emph{different} operational scenario from that of Ozawa. Rather than
asking whether we can beat the SQL by specific measure-and-reprepare protocols
on a single particle, with the aid of an \emph{external} reference frame, this
paper addresses the intrinsic limitations stemming from a composite quantum
system undergoing free evolution and serving as the only resource to
define the space and time coordinates of another system. As we show in
Sec.~\ref{sec:internal-perspective}, our findings are relevant in the context
of QRFs and relational observables, which are measured without assuming access
to an external reference frame. 

Furthermore, we observe that the measurement protocol discussed above focuses on
position measurements of a \emph{single} quantum particle, while our interest
ultimately lies in joint measurements of \emph{position and time} of a
\emph{composite} quantum system. In particular, in
Sec.~\ref{sec:external-perspective} we analyse how the clock's temporal
precision depends on clock coherence in the internal-energy basis, and how this
in turn depends on the spread in momentum observables (see
Eqs.~\eqref{eq:clock-off-diag-terms}, \eqref{eq:QSL-qubit-static}). Although contractive states can reach a small position uncertainty after a definite time $\tau_c$, they do so at the expense of a large momentum uncertainty.
Explicitly, contractive Gaussians in phase space can be written as 
\begin{equation}\label{eq_app:contractive-states}
    \psi(x) = \frac{1}{(2\pi\sigma^2)^{1/4}}
    \exp\!\left[-\frac{1 - 2 i \gamma}{4\sigma^2}\,x^2\right],
\end{equation}
for which $\Delta^2_\psi\, \hat{x}(0) = \sigma^2$ and
$\Delta^2_\psi\, \hat{p} = \frac{\hbar^2}{4\sigma^2}\,(1 + 4\gamma^2)$. Thus, a strong
contraction (small $\Delta_\psi\, \hat{x}_{\min}$) requires $|\gamma|\gg 1$, which in turn
implies a large momentum spread, $\Delta_\psi\, \hat{p} \propto |\gamma|$, i.e.\ the
protocol prepares states with significantly increased $\Delta_\psi\,\hat{p}$. In light of our analysis in Sec.~\ref{sec:external-perspective}, this increase in momentum uncertainty would inevitably reduce the temporal precision of the clock.

Although it would be interesting to see whether Ozawa’s argument can be
extended to our scenario (in particular in the framework of QRFs and relational
observables), in the context of this work it is more natural to consider
localized states which are symmetric in phase space, i.e.\ with
$\mathrm{Cov}_\Psi(\hat x,\hat v) = 0$.\footnote{We could simply assume positive correlations, however the effect of a positive covariance term is not very relevant in the context of our work. To see this, we first note that correlations (Eq. \eqref{eq_app:contractive-states}), whether positive or negative, require a large momentum spread and thus worsen clock precision; secondly, we do not consider protocols that would actively create such correlations by means of an external frame.}

\section{Minimum Uncertainty States (MUS)}\label{app:MUS-states}

\noindent The Salecker–Wigner thought experiment yields the minimum uncertainty in position for a free single particle (Eq. \eqref{eq:Salecker-Wigner-bound}). The optimal quantum states are those that minimize the uncertainty relation between position and momentum:

\begin{equation}\label{eq:Heis-Unc-Rel-XP}
    \Delta^2_\Psi \hat{x}\,\Delta^2_\Psi \hat{p} \;\ge\; \frac{\hbar^2}{4},
\end{equation}
i.e., the standard Gaussian states in phase space.

In Sec. \ref{sec:external-perspective} we show that, for a free \emph{composite} particle, the minimum uncertainty in position is different, since it is modified by the internal degrees of freedom -- the clock. From our derivation it is clear that the optimal states are not the usual Gaussian in phase space but those states minimizing the uncertainty relation between position and velocity operators:

\begin{equation}\label{appeq:Heis-Unc-Rel-XV}
  \Delta_\Psi^2 \hat{x}\;\Delta_\Psi^2 \hat{v}
   \;\ge\; \frac{\hbar^2}{4}\,\big\langle \hat{m}^{-1}\big\rangle_\Psi^2 .
\end{equation}
A full derivation and a detailed analysis of this class of states is given in \cite{Zych-MUS-states}, where they are referred to as ``minimum-uncertainty-states" (MUS). In general, they are entangled states of the form:

\begin{equation}\label{eqapp:MUS-C-state}
   \ket{\Psi} \;=\; \sum_i \phi_i \int_{\mathbb{R}} dp\; \psi_{v_0,x_0}(p,\epsilon_i)\,\ket{p}_{r}\otimes\ket{\epsilon_i}_{c},
\end{equation}
In particular, we focus on symmetric wavepackets ($\mathrm{Cov}_\Psi(\hat{x}, \hat{v})=0$), which read:
\begin{equation}\label{eqapp:C-Gaussian-wavepacket}
   \psi_{v_0,x_0}(p,\epsilon_i) = \mathcal{N}_i
   \exp\!\left[-\frac{(p-m_i v_0)^2}{2m_i\hbar\Omega}-\frac{i}{\hbar}(p-m_i v_0)x_0\right]
\end{equation}
where $\mathcal{N}_i = \left(\pi\hbar m_i\Omega\right)^{-1/4}$ is the
normalization constant and $\Omega$ fixes the scale of the spread. Note
that both the average momentum and its variance depend on the internal energy through $m_i=m+\epsilon_i/c^2$, while the
average velocity $v_0$ is the same across the internal-energy branches.

By contrast, the states that saturate \eqref{eq:Heis-Unc-Rel-XP} are separable,
\begin{equation}\label{eq:MUS-P-state}
   \ket{\Psi} \;=\; \left(\int_{\mathbb{R}} dp\, \Psi_{p_0,x_0}(p)\,\ket{p}_{r}\right) \;\otimes\; \left(\sum_i \phi_i \ket{\epsilon_i}_{c}\right),
\end{equation}
with Gaussian wavepackets
\begin{equation}\label{eq:P-Gaussian-wavepacket}
  \Psi_{p_0,x_0}(p)
  = \left(\frac{2\sigma^2}{\pi \hbar^2}\right)^{1/4}
    \exp\!\left(-\,\frac{(p-p_0)^2\sigma^2}{\hbar^2}
                -\frac{i}{\hbar}(p-p_0)x_0\right),
\end{equation}
All internal branches share the same peak momentum $p_0$ but have branch-dependent velocities $v_{0,i}=p_0/m_i$, which induces a drift in configuration space. At early times the composite particle is localized within $\Delta \hat{x}(0)=\sigma$. During free evolution, each internal-energy branch (with mass $m_i$) propagates with a different velocity $v_{0,i}=p_0/m_i$, contributing an additive term to the spread in position space. Interestingly, even in the frame where $p_0=0$, Gaussian states in phase-space remain suboptimal for spatial localization, as we show below.\\

\paragraph*{Gaussian states in phase space}

We consider the standard Gaussian states \eqref{eq:MUS-P-state} and we denote the related quantities with the superscript $G$. The spread in position (Eq. \eqref{eq:X-variance-free} in the main text) reads:

\begin{gather}\label{eq:X-variance-P-full}
    \Delta_\Psi^2 \hat{x}(t)^G \;=\; \sigma^2 \;+\; \frac{\hbar^2 t^2}{4\sigma^2}\,\big\langle \hat{m}^{-2}\big\rangle_\Psi
    \;+\; p_0^2 t^2\,\Delta^2_\Psi\!\big(\hat{m}^{-1}\big).
\end{gather}
Physically, the first term is the standard Gaussian spreading, averaged over the superposed mass-energies. The second term describes the delocalization due to the internal-energy branches travelling at different (average) velocities $v_{0,i} = p_0/m_i$. Minimizing \eqref{eq:X-variance-P-full} with respect to $\sigma$ yields:

\begin{eqnarray}\label{eq:X-variance-P-minimized}
     \Delta_\Psi^2 \hat{x}(t)^G \,\geq\, p_0^2 t^2\,\Delta^2_\Psi\!\big(\hat{m}^{-1}\big) \;+\; \hbar |t|\,\sqrt{\big\langle \hat{m}^{-2}\big\rangle_\Psi}.
\end{eqnarray}
Proceeding as in the main text (see Eq. \eqref{eq:X-variance-free-avg-energy}) we find

\begin{gather}\label{eq:X-variance-P-final}
    \Delta_\Psi^2 \hat{x}(t)^G \,\geq\, \left(\frac{p_0 t}{m}\right)^2 \frac{\Delta_\Psi^2 \hat{H}_c}{m^2 c^4}
    \;+\; \frac{\hbar|t|}{\overline{m}}+\frac{3}{2}\frac{\hbar|t|}{m}\frac{\Delta_\Psi^2 \hat{H}_c}{m^2 c^4}.
\end{gather}
Eq. \eqref{eq:X-variance-P-final} shows two contributions from the internal-energy spread: (i) a kinematical contribution $\propto p_0^2t^2$ (ii) an enhancement of the Gaussian spreading. Comparing the latter with the general bound of Eq. \eqref{eq:X-variance-free-final}, we see that it has a strictly larger prefactor: \(3/2\) here versus \(1\) in Eq. \eqref{eq:X-variance-free-final}. Even when $p_0=0$, phase-space MUS are therefore not maximally localized in space-time.\\

\paragraph*{MUS states}

Now we repeat the analysis for the MUS states of Eq. \eqref{eqapp:MUS-C-state} and we denote the related quantities with the superscript $M$. Computing the initial spreads:

\begin{eqnarray}\label{eq:MUS-C-variances}
   \Delta^2_\Psi \hat{x}(0)^M=\Big\langle \frac{\hbar}{2\hat{m}\Omega}\Big\rangle_\Psi,
   \qquad
   \Delta_\Psi^2 \hat{v}^M=\Big\langle \frac{\hbar\Omega}{2\hat{m}}\Big\rangle_\Psi.
\end{eqnarray}
we find the uncertainty in position \eqref{eq:X-variance-free}:

\begin{eqnarray}\label{eq:X-variance-C-full}
   \Delta_\Psi^2 \hat{x}(t)^M=\Big\langle \frac{\hbar}{2\hat{m}\Omega}\big(1+\Omega^2 t^2\big)\Big\rangle_\Psi,
\end{eqnarray}
Notice that, as expected, the kinematical term $\sim v_0$ is absent. Minimizing Eq. \eqref{eq:X-variance-C-full} with respect to $\Omega$ gives

\begin{eqnarray}\label{eq_app:X-variance-C-minimized}
   \Delta_\Psi^2 \hat{x}(t)^M\,\geq\,\hbar|t|\;\big\langle \hat{m}^{-1}\big\rangle_\Psi,
\end{eqnarray}
which coincides with the general minimum in Eq. \eqref{eq:variance-X-free-minimized}. These states are therefore the most localized in space at fixed $\Delta_\Psi \hat{H}_c$. 
Interestingly, the same entanglement with the c.o.m. degrees of freedom that makes these states the most localized in space also reduces internal coherence, making them less precise as clocks. We refer the reader to the Appendix \ref{app:trade-off} for a detailed discussion and a quantitative analysis.

\section{Temporal uncertainty of mixed-state clocks}\label{app:clock-precision}

\noindent In this appendix, we study how the coupling between the internal and c.o.m. degrees of freedom of a composite particle affects its function as a quantum clock.  
In particular, we show how relativistic mass–energy coupling leads to decoherence in the internal-energy basis, thereby reducing the clock’s temporal precision. Finally, we compare this effect for the two classes of states analysed in App. \ref{app:MUS-states}, showing that the states that are more localized in space, because this requires entanglement between the clock and the c.o.m., are less precise as clocks.

\subsection{Decoherence}

Because the Hamiltonian of a composite particle (Eq. ~\eqref{eq:H_SR_exact}) couples its internal and c.o.m.\ degrees of freedom, the internal sector undergoes decoherence during free evolution. If, in addition, the initial state is entangled—as for instance in the MUS states of Eq.~\eqref{eqapp:MUS-C-state}—the clock’s internal state is already mixed at $t=0$. 

Consider a discrete clock with $ \hat{H}_c = \sum_i\, \epsilon_i \, \ketbra{\epsilon_i}{\epsilon_i}_c.$ In the internal-energy basis, the clock’s state reads $ \hat{\rho}_c(t) = \sum_{i,j} \hat{\rho}_c^{ij}(t)\,  \ketbra{\epsilon_i}{\epsilon_j}_c$. The off-diagonal elements $|\hat{\rho}_c^{ij}(t)|$ (with $i\neq j$) quantify the internal coherence of the clock.

We consider pure STQRF states of the general form
\begin{gather}\label{eqapp:general-state}
    \ket{\Psi(t)} = \sum_i \,  \phi_i(t)\,\!\int_{\mathbb{R}}\! dp\, \psi_i(p,t)\, \ket{p}_r \!\otimes\! \ket{\epsilon_i}_c,
\end{gather}
where $\psi_i(p,t)$ is the c.o.m.\ wave packet that in general depends on the internal level $\epsilon_i$ (through the mass $m_i = m + \epsilon_i/c^2$), and $\phi_i=\sqrt{p_i}$ is the probability amplitude in the internal-energy basis. Tracing over the c.o.m.\ degrees of freedom gives
\begin{equation}
    \hat{\rho}_c^{ij}(t) = \phi_i(t)\phi_j^*(t)
    \int_{\mathbb{R}} dp\, \psi_i(p,t)\psi_j^*(p,t),
\end{equation}
where the integral represents the overlap between momentum-space wave packets associated with internal energies $\epsilon_i$ and $\epsilon_j$. Time evolution is generated by the total Hamiltonian in Eq. ~\eqref{eq:H_SR_exact} with eigenvalues
\(
E_i(p) = p^2/2m_i + m_i c^2
\)
and $m_i = m + \epsilon_i / c^2$. Hence,
\begin{gather}\label{eq_appC:clock-state-time-evolution}
    \hat{\rho}_c^{ij}(t)
    = \hat{\rho}_c^{ij}(0)\,
    e^{-\,\frac{i(m_i - m_j)c^2 t}{\hbar}}\,
    \big\langle \big( e^{-i\,\alpha_{ij}(t)\,\hat{p}^2} \big)_{ij}\, \rangle_r,
\end{gather}
where
\begin{equation}
\alpha_{ij}(t) := \frac{t}{2\hbar}\!\left(\frac{1}{m_i} - \frac{1}{m_j}\right),
\end{equation}
and we defined the normalized cross average with respect to the overlap $\hat{\rho}_c^{ij}(0) = \displaystyle\int_{\mathbb{R}} dp\; \bra{p} \hat{\rho}_{rc}^{ij} \ket{p}_r $:

\begin{gather}\label{eqapp:SQL-cross-avg}
    \langle \big(f(\hat{p})\big)_{ij}\rangle_r:=\frac{\int_{\mathbb{R}} dp\,\bra{p}\hat{\rho}_{rc}^{ij}\ket{p}_r\, f(p)}{\int_{\mathbb{R}} dp\,\bra{p}\hat{\rho}_{rc}^{ij}\ket{p}_r},
\end{gather}
and $\hat{\rho}_{rc}^{ij} = \bra{\epsilon_i}\hat{\rho}_{rc}\ket{\epsilon_j}_c$.\\

\paragraph*{Low-energy expansion.}
Expanding $m_i=m+\epsilon_i/c^2$ to first order gives
\begin{equation}
    \alpha_{ij}(t)\;=\;\frac{t}{2\hbar m}\frac{\Delta\epsilon_{ij}}{m c^2} + o\left(\left(\tfrac{\Delta\epsilon_{ij}}{m c^2}\right)^2\right),
\end{equation}
where \(\Delta\epsilon_{ij}\) is the level splitting. For small $\alpha_{ij}$, we can expand the exponential up to second order $o\,\Big(\big(\Delta \epsilon_{ij}/mc^2\big)^{\!2}\Big)$:
\begin{equation}
\Big|\big\langle \left(e^{i\alpha_{ij}(t)\hat{p}^2}\right)_{ij}\;\big\rangle_r\Big|
=
1-\frac{1}{2}\alpha_{ij}^2(t)\,\Delta^2_r(\hat{p}^2_{ij})
+O(\alpha_{ij}^3),
\end{equation}
where
\(
\Delta^2_r(\hat{p}^2_{ij})
=\langle \hat{p}^4_{ij}\rangle_r- \big( \langle \hat{p}^2_{ij}\rangle_r \big)^2
\). Combining the above relations, we obtain the short-time decay of the off-diagonal coherence up to order $o\,\Big(\big(\Delta \epsilon_{ij}/mc^2\big)^{\!2}\Big)$:

\begin{gather}
|\hat{\rho}_c^{\,ij}(t)|
= |\hat{\rho}_c^{\,ij}(0)|
\Bigg[1-\frac{1}{2}
\Bigg(\frac{t}{2\hbar m }\frac{\Delta \epsilon_{ij}}{mc^2}\Bigg)^{\!2}
\Delta^2_r(\hat{p}^2_{ij})\Bigg]
\label{eqapp:rho-ij-final}
\end{gather}
This shows that decoherence of the clock’s internal state is governed by (i)~initial entanglement, through \(|\hat{\rho}_c^{\,ij}(0)| \leq1\); and  (ii)~dynamical entanglement induced by quantum uncertainty in the SR time dilation factor  
\(\Delta(\hat{p})=1-\frac{\hat{p}^2}{2m^2c^2}\), which yields the time-dependent term.

\subsection{Quantum speed limit}\label{app:C2-QSL}

The QSL found in \cite{QSL-coherence-Spekkens} assumes that the clock's state undergoes a unitary dynamics, i.e., $\rho_c(t) = e^{-\frac{i}{\hbar}t\hat{H}_c}\rho e^{+\frac{i}{\hbar}t\hat{H}_c}$, in which case the coherence is constant, fixed by the initial state.

Here, we extend it to our case, where the reduced clock’s state does not undergo a simple unitary dynamics, rather a ``random-unitary dephasing". Explicitly, considering the first-order Hamiltonian of Eq. \eqref{eq:H_SR_expanded},

\begin{gather}\label{eq_appC:clock-mixture-unitaries}
    \hat{\rho}_c(t)
    = \mathrm{tr}_r[\hat{\rho}_{rc}(t)]
    = \int_{\mathbb{R}} dp\, e^{-\frac{i}{\hbar} t\,\Delta(p) \hat{H}_c}\,
    \bra{p}\hat{\rho}_{rc}\ket{p}_r\,
    e^{\frac{i}{\hbar} t\,\Delta(p) \hat{H}_c}.
\end{gather}

where $\Delta(p)=1-p^2/2m^2c^2$ is the time dilation factor. Following \cite{QSL-coherence-Spekkens}, we have

\begin{gather}
    \|\hat{\rho}_c(t)-\hat{\rho}_c\|_1 = \Big\|\int_0^t ds \; \frac{d}{ds}\hat{\rho}_c(s)\Big\|_1 \leq \int_0^t ds \,\|  \dot{\hat{\rho}}_c(s)\|_1\, ,
    \label{eq_appC:clock-trace-distance}
\end{gather}
where $\|\hat{O} \|_1 \coloneq \mathrm{tr}\!\Big[\sqrt{\hat{O}^\dagger \hat{O}}\,\,\Big]$ is the trace norm. In general, the distance is expressed in terms of the rate of evolution of the state $\hat{\rho}_c$. Therefore, we expect that the special relativistic time dilation factor will play a role. From Eq. \eqref{eq_appC:clock-mixture-unitaries}, 

\begin{gather}
    \dot{\hat{\rho}}_c(s) = \frac{i}{\hbar}\int_{\mathbb{R}} dp \;\Delta(p)\; \bigg[ \bra{p} \hat{\rho}_{rc}(s) \ket{p}_r,\hat{H}_c \bigg]
\end{gather}

Hence, Eq. \eqref{eq_appC:clock-trace-distance} reads:

\begin{gather}
    \|\hat{\rho}_c(t)-\hat{\rho}_c\|_1 \leq \int_0^t \frac{ds}{\hbar}\Big\|\int_{\mathbb{R}} dp \;\Delta(p)\;  \big[ \bra{p} \hat{\rho}_{rc}(s) \ket{p}_r,\hat{H}_c \big]\Big\|_1 \nonumber \\ 
    \leq \int_0^t \frac{ds}{\hbar}\int_{\mathbb{R}} dp \;\Delta(p)\; \Big\| \big[ \bra{p} \hat{\rho}_{rc} \ket{p}_r,\hat{H}_c \big]\Big\|_1 \label{eq_appC:extended_bound}
\end{gather}
where the last line follows by using the triangle inequality and the invariance of the trace norm under unitaries. The first inequality is more general and takes into account the dynamical decoherence, while the second one takes into account only the entanglement of the initial state.

\subsubsection{Static QSL}

From the second line of Eq. \eqref{eq_appC:extended_bound}, defining $\delta \coloneq \|\hat{\rho}_c(t_\delta)-\hat{\rho}_c\|_1$, we immediately find a QSL in the form

\begin{gather}\label{eq_appC:static-QSL}
  t_{\delta} \geq \frac{\hbar\delta}{ F_{\hat{H}_c}(\hat{\rho}_c)}
\end{gather}
where

\begin{equation}\label{eq_appC:static-QSL-F}
     F_{\hat{H}_c}(\hat{\rho}_c)
     \coloneq \int_{\mathbb{R}} dp\;\Delta(p)\;
     \big\|\,[\bra{p}\hat{\rho}_{rc}\ket{p}_r,\hat{H}_c]\,\big\|_1,
\end{equation}
We write the commutator in the energy basis

\begin{gather}
    \bigg[\bra{p} \hat{\rho}_{rc} \ket{p}_r,\hat{H}_c\bigg] = \sum_{ij} (\epsilon_j - \epsilon_i)\,\bra{p} \hat{\rho}_{rc}^{ij}\ket{p}_r\, \ketbra{\epsilon_i}{\epsilon_j}_c
    \label{eq_appC:commutator-energy-basis}
\end{gather}
so that Eq. \eqref{eq_appC:static-QSL-F} becomes:

\begin{gather}
     F_{\hat{H}_c}(\hat{\rho}_c)
     = \sum_{ij}  \big|\epsilon_j - \epsilon_i\big| \, \big|\hat{\rho}_c^{ij}(0)\big|\, \;\Big|\langle \big(\Delta(\hat{p})\big)_{ij}\;\rangle_r\Big| \nonumber\\
     = \sum_{ij}  \big|\epsilon_j - \epsilon_i\big| \, \big|\hat{\rho}_c^{ij}(0)\big|\, \;\Biggl(1\;-\,\frac{\langle \hat{p}^2_{ij}\rangle_r }{2m^2c^2} \Biggr)
     \label{eq_appC:static-QSL-F-explicit}
\end{gather}
This equation shows that temporal precision is decreased not only because of the entanglement of the initial state, but also because of special relativistic time dilation, which slows down the rate-of-evolution of the clock's state.

In the following, we explicitly compute the SQL (Eq. \eqref{eq_appC:static-QSL-F-explicit}) for the two classes of states analysed in App. \ref{app:MUS-states}. 

We'll consider the simplest example of a qubit clock, in which case the dependence on the energy spread $\Delta_\Psi \hat{H}_c$ is manifest. Alternatively, one could reformulate the SQL in terms of $2$-norms by using the inequality $\|O\|_1 \leq \sqrt{\mathrm{rg}(O)} \,\|O\|_2$. In this case, the energy spread comes out by noting that $\sum_{i\neq j} (\epsilon_i - \epsilon_j)^2 \; p_i\, p_j = 2\Delta^2_{\Psi}H_c$. Thus, the results that we find for a 2-state clock, can be straightforwardly generalized to any dimension. \\

\paragraph*{Gaussian in phase-space (separable).}
We consider the standard Gaussian wavepackets in phase space (Eq.~\eqref{eq:MUS-P-state}) and we denote the related quantities with the superscript $G$. The c.o.m. wave-packets are independent of the internal energy, so $\big|\hat{\rho}_c^{ij}(0)\big|=1$ and $\langle f(\hat{p})\rangle_{ij} \equiv \langle f(\hat{p})\rangle_\Psi$. Thus, we find:

\begin{gather}\label{eq_app:Vt-C}
    F^G_{\hat{H}_c}(\hat{\rho}_c) = \sum_{ij} \sqrt{p_i\,p_j}\;\big|\epsilon_j - \epsilon_i\big|\, \Bigg(1-\frac{\sigma^2 + p_0^2}{2m^2c^2}\Bigg).
\end{gather}

For simplicity, let's consider a qubit clock with $p_i=p_j = 1/2$ and $\Delta\epsilon = 2\Delta_{\Psi}\hat{H}_c$. The QSL of Eq. \eqref{eq_appC:static-QSL} reads

\begin{gather}\label{eq_appC:QSL-G-static}
    t^G_\delta \geq \frac{\delta \hbar}{\Delta_\Psi \hat{H}_c} \Bigg(1+\frac{\Delta^2_\Psi \hat{p} + p_0^2}{2m^2c^2}\Bigg)
\end{gather}\\

\paragraph*{MUS (entangled).}
We consider the MUS in configuration space (Eq.~\eqref{eq_app:C-Gaussian}) and we denote the related quantities with the superscript $M$. The c.o.m. wave-packets depend on the internal energy:

\begin{equation}\label{eq_app:C-Gaussian}
   \psi_{v_0,x_0=0}(p,\epsilon_i)=\left(\pi\hbar m_i\Omega\right)^{-1/4}
  \exp\Big({-\frac{(p-m_i v_0)^2}{2m_i\hbar\Omega}}\Big),
\end{equation}
The product between two of them is another Gaussian with $\sigma_{ij}^2=\frac{m_i m_j}{m_i+m_j}\,\hbar\Omega$ and $\mu_{ij}=\frac{2m_i m_j}{m_i+m_j}\,v_0$. Hence: 

\begin{align}
    \big|\hat{\rho}_c^{ij}(0)\big|^M
= \sqrt{p_ip_j}\sqrt{\frac{2\sqrt{m_i m_j}}{m_i+m_j}}\,
\exp\!\left[-\frac{(m_i-m_j)^2 v_0^2}{2\hbar\Omega(m_i+m_j)}\right].
\end{align}
Now, we Taylor-expand up to second order $o\,\Big(\big(\Delta \epsilon_{ij}/mc^2\big)^{\!2}\Big)$, which gives:

\begin{align}
\big|\hat{\rho}_c^{ij}(0)\big|^M
= \sqrt{p_ip_j} \Bigg( 1-\left(\frac{\Delta\epsilon_{ij}}{mc^2}\right)^2\!\left(\frac{1}{16}+\frac{m v_0^2}{4\hbar\Omega}\right)\Bigg).
\label{eq_app:initial-visibility-qubit-C}
\end{align}
Notice that the initial entanglement manifests in two terms:
(i) a constant term due to the Gaussian widths being dependent on the internal energy, $\Delta^2_{r} (\hat{p}_{ii})= \hbar\Omega m_i /2$, and (ii) a frame-dependent term $\sim v_0$, due to the average momenta being dependent on the internal energy, $p_0 = v_0m_i$. Similarly, because  we have

\begin{gather}
    \frac{m_i m_j}{m_i+m_j} = \frac{ m}{2}\left(1+ \frac{\epsilon_{i}+\epsilon_j}{mc^2} - \frac{\bigl( \epsilon_i-\epsilon_j\bigr)^2}{4m^2c^4}\right) + o\,\Big(\big(\tfrac{\Delta \epsilon_{ij}}{mc^2}\big)^{\!3}\Big)
\end{gather}
we find

\begin{gather}
    \frac{\langle \hat{p}_{ij}^2\rangle}{2m^2c^2} = \frac{\frac{\hbar\Omega}{2}\overline{m}_{ij}+\overline{m}_{ij}^2v_0^2}{2m^2c^2} + o\,\Big(\big(\tfrac{\Delta \epsilon_{ij}}{mc^2}\big)^{\!3}\Big)
\end{gather}
where $\overline{m}_{ij} = m+(\epsilon_i+\epsilon_j)/2c^2$. Combining Eqs. \eqref{eq_app:initial-visibility-qubit-C}–\eqref{eq_app:visibility-qubit-C-variance-p^2} gives, up to $o\,\Big(\big(\Delta \epsilon_{ij}/mc^2\big)^{\!2}\Big)$),
\begin{gather}\label{eq_app:Vt-C-qubit}
    F^M_{\hat{H}_c}(\hat{\rho}_c) = \sum_{ij} \sqrt{p_ip_j}\, \big|\epsilon_j - \epsilon_i\big|\, \Bigg(1-\frac{\frac{\hbar\Omega}{2}\overline{m}_{ij}+\overline{m}_{ij}^2v_0^2}{2m^2c^2}\nonumber \\
    -\left(\frac{\Delta\epsilon_{ij}}{mc^2}\right)^2\!\left(\frac{1}{16}+\frac{m v_0^2}{4\hbar\Omega}\right)\Bigg).
\end{gather}

Considering a qubit clock with $p_i=p_j = 1/2$ and $\Delta\epsilon = 2\Delta_{\Psi}\hat{H}_c$, the QSL of Eq. \eqref{eq_appC:static-QSL} reads

\begin{gather}
    t^M_\delta \geq \frac{\delta \hbar}{\Delta_\Psi \hat{H}_c} \Bigg(1+\frac{\frac{\hbar\Omega}{2}\overline{m}+\overline{m}^2v_0^2}{2m^2c^2}\nonumber \\
    +\frac{\Delta^2_\Psi \hat{H}_c}{m^2c^4}\!\left(\frac{1}{4}+\frac{m v_0^2}{\hbar\Omega}\right)\Bigg)
\end{gather}

Neglecting terms of order
$\sim \big\langle(\hat{p}^2 /m^2c^2)(\hat{H}_c/mc^2)\big \rangle_\Psi$, allows us to identify
the average momentum and its spread as $m v_0 \equiv p_0$ and
$\hbar\Omega m/2 \equiv \Delta^2_\Psi \hat{p}$. Thus, Eq. \eqref{eqapp:QSL-dynamical-c} becomes

\begin{gather}
    t^M_\delta \geq \frac{\delta \hbar}{\Delta_\Psi \hat{H}_c} \Bigg(1+\frac{\Delta^2_\Psi \hat{p} + p_0^2}{2m^2c^2} + \nonumber \\
    \frac{\Delta^2_\Psi \hat{H}_c}{m^2c^4}\!\left(\frac{1}{4}+\frac{p_0^2}{2\Delta^2_\Psi \hat{p}}\right)
    \Bigg)\label{eq_appC:QSL-M-static}
\end{gather}

Comparing this QSL to that of standard Gaussian states in Eq. \eqref{eq_appC:QSL-G-static}, we see that MUS have a larger temporal uncertainty precisely because of the initial entanglement (last term of \eqref{eq_appC:QSL-M-static}).

\subsubsection{Dynamical QSL}

To take into account the dynamical decoherence, we consider the first inequality of Eq. \eqref{eq_appC:extended_bound} that, combined with Eqs. \eqref{eq_appC:clock-state-time-evolution} and \eqref{eq_appC:commutator-energy-basis}, gives:

\begin{gather}
    \delta 
    \leq \;\sum_{ij}  \big|\epsilon_j - \epsilon_i\big| \, \int_0^{t_\delta} \frac{ds}{\hbar}\Big|\biggl( \int_{\mathbb{R}} dp \;\Delta(p)\;  \,\bra{p} \hat{\rho}_{rc}^{ij}(s) \ket{p}_r \biggr)\Big|\,\nonumber \\
    = \sum_{ij}  \big|\epsilon_j - \epsilon_i\big| \, \big|\hat{\rho}_c^{ij}(0)\big|\,\int_0^{t_\delta} \frac{ds}{\hbar} \;\Bigg|\Big\langle \, \Big( \Delta(\hat{p}) \,e^{i\alpha_{ij}(s)\hat{p}^2}\Big)_{ij}\;\Big\rangle_r\Bigg|
\end{gather}
Now we expand up to $o\,\Big(\big(\Delta \epsilon_{ij}/mc^2\big)^{\!2}\Big)$:

\begin{gather}
    \;\Bigg|\Big\langle \, \Big( \Delta(\hat{p}) \,e^{i\alpha_{ij}(s)\hat{p}^2}\Big)_{ij}\;\Big\rangle_r\Bigg| = 
\Biggl(1\; -\,\frac{\langle \hat{p}^2_{ij}\rangle_r}{2m^2c^2}
\;\nonumber \\-\;\frac{1}{2}\,\biggr(\frac{s}{2\hbar m}\,\frac{\Delta\epsilon_{ij}}{m c^2}\biggl)^2\,\Delta^2_r(\hat{p}^2_{ij})\biggr)
    \label{eq_appC:dynamical-QSL-integral}
\end{gather}
The first term is the \emph{static} reduction due to the average time dilation factor, that we also have in Eq. \eqref{eq_appC:static-QSL-F-explicit}; the third term is the \emph{dynamical} decoherence governed by the uncertainty in the time dilation operator $\Delta(\hat{p})$ through \(\Delta^2_r(\hat{p}^2_{ij})\). Integrating Eq. \eqref{eq_appC:dynamical-QSL-integral} over time yields a cubic equation:

\begin{gather}
\delta \;\lesssim\;t_\delta
\sum_{i, j}\frac{|\Delta\epsilon_{ij}|}{\hbar}\, \big|\hat{\rho}_c^{ij}(0)\big|\, \Biggl(1\;-\,\frac{\langle \hat{p}^2_{ij}\rangle_r}{2m^2c^2} \nonumber \\
\;-\;\frac{1}{6}\,\biggr(\frac{t_\delta}{2\hbar m}\,\frac{\Delta\epsilon_{ij}}{m c^2}\biggl)^2\,\Delta^2_r(\hat{p}^2_{ij})\biggr).
\end{gather}
Defining the ``static" and ``dynamical" contributions

\begin{gather}
    S := \sum_{i\ne j}|\Delta\epsilon_{ij}|\,\big|\hat{\rho}_c^{ij}(0)\big|\,
    \Big(1-\frac{\langle \hat{p}^2_{ij}\rangle_r}{2m^2c^2}\Big),\nonumber \\
    D := \sum_{i\ne j}|\Delta\epsilon_{ij}|\,\big|\hat{\rho}_c^{ij}(0)\big|\,
    \Delta^2_r(\hat{p}^2_{ij})\left(\frac{\Delta\epsilon_{ij}}{2\hbar m^2 c^2}\right)^2,
\end{gather}
we can find a QSL for small times by solving the cubic equation perturbatively. At first order, we find:

\begin{equation}
t_\delta \;\gtrsim\;
\frac{\delta \hbar}{S}\left(
1+\frac{\hbar^2 D}{6 S^{\,3}}\,\delta^{2}
\right)
\label{eq_app:QSL_refined}
\end{equation}
For a qubit clock with $\Delta\epsilon = 2\Delta_{\Psi}\hat{H}_c$ it reads:

\begin{gather}
    t_\delta \geq \frac{\delta\hbar}{2\Delta_{\Psi}\hat{H}_c} \frac{1}{\big|\hat{\rho}_c^{01}(0)\big|} \Bigg( 1+\,\frac{\langle \hat{p}^2_{ij}\rangle_r}{2m^2c^2} \nonumber \\
\;+\;\frac{\delta^2}{6}\,\frac{\Delta^2_r(\hat{p}^2_{ij})}{4m^4c^4} \,\big|\hat{\rho}_c^{01}(0)\big|^{-2}\Bigg)
\label{eqapp:QSL-dynamical-qubit}
\end{gather}
Last term in Eq. \eqref{eqapp:QSL-dynamical-qubit} is the loss in time precision due to the dynamical decoherence (see Eq. \eqref{eqapp:rho-ij-final}. In the following, we compare the two classes of states analysed in App. \ref{app:MUS-states}.\\

\paragraph*{Phase-space MUS (separable).}

For the standard Gaussians in phase space (Eq.~\eqref{eq:MUS-P-state}), 

\begin{equation}\label{eq_app:visibility-qubit-P-variance-p^2}
    \Delta^2_r(\hat{p}^2_{ij})^G \equiv  \Delta^2_{\Psi}(\hat{p}^2)=2\Delta^2_\Psi \hat{p}\left(\Delta^2_\Psi \hat{p} + 2p_0^2\right).
\end{equation}
For a qubit clock with $p_i=p_j = 1/2$ the QSL \eqref{eqapp:QSL-dynamical-qubit} reads:

\begin{gather}
      t^G_\delta \geq \frac{\delta \hbar}{\Delta_\Psi \hat{H}_c} \Bigg(1+\frac{\Delta^2_\Psi \hat{p} + p_0^2}{2m^2c^2}+\frac{\delta^2}{6} \Big( \frac{2\Delta^2_\Psi \hat{p}}{m^2c^2} \Big)^2 \left(\frac{1}{4}+\frac{p_0^2}{2\Delta^2_\Psi \hat{p}}\right)\Bigg)
      \label{eqapp:QSL-dynamical-P}
\end{gather}\\

\paragraph*{Configuration-space MUS (entangled).}
For the MUS in configuration space (Eq.~\eqref{eq_app:C-Gaussian}), we have (using the Gaussian identity $ \mathrm{Var}(X^2) = 2\mathrm{Var}(X)^4 + 4\left<X\right>^2 \mathrm{Var}(X)^2$)

\begin{equation}\label{eq_app:visibility-qubit-C-variance-p^2}
    \Delta^2_r(\hat{p}^2_{ij})^M = \frac{(\hbar m\Omega)^2}{2}+2\,\hbar\Omega\,m^3 v_0^2+o\left(\frac{\Delta\epsilon_{ij}}{mc^2}\right).
\end{equation}
For a qubit clock with $p_i=p_j = 1/2$ the QSL \eqref{eqapp:QSL-dynamical-qubit} reads:

\begin{gather}
    t^M_\delta \geq \frac{\delta \hbar}{\Delta_\Psi \hat{H}_c} \Bigg(1+\frac{\frac{\hbar\Omega}{2}\overline{m}+\overline{m}^2v_0^2}{2m^2c^2}\nonumber \\
    + \frac{\delta^2}{3} \Big( \frac{\hbar\Omega}{mc^2} \Big)^2 \left(\frac{1}{4}+\frac{m v_0^2}{\hbar\Omega}\right)+\frac{\Delta^2_\Psi \hat{H}_c}{m^2c^4}\!\left(\frac{1}{4}+\frac{m v_0^2}{\hbar\Omega}\right) \Bigg)
    \label{eqapp:QSL-dynamical-c}
\end{gather}

Neglecting terms of order
$\sim \big\langle(\hat{p}^2 /m^2c^2)(\hat{H}_c/mc^2)\big \rangle_\Psi$, allows us to identify
the average momentum and its spread as $m v_0 \equiv p_0$ and
$\hbar\Omega m/2 \equiv \Delta^2_\Psi \hat{p}$. Thus, Eq. \eqref{eqapp:QSL-dynamical-c} becomes

\begin{gather}
  t^{M}_{\delta}
  \geq
  \left(\frac{\delta \hbar}{\Delta_\Psi \hat{H}_c}\right)
  \Bigg[
  1+\frac{\Delta^2_\Psi \hat{p} + p_0^2}{2m^2c^2}\nonumber \\
  +\frac{\delta^2}{3}\left(\frac{2\Delta^2_\Psi \hat{p}}{m^2c^2}\right)^{\!2}
  \left(\frac{1}{4}+\frac{p_0^2}{2\Delta^2_\Psi \hat{p}}\right)
  +\frac{\Delta_\Psi^2 \hat{H}_c}{m^2c^4}\left(\frac{1}{4}+\frac{p_0^2}{2\Delta^2_\Psi \hat{p}}\right)
  \Bigg].
  \label{eq_appC:QSL-M-dynamical}
\end{gather}

Comparing this QSL to that of standard Gaussian states in Eq. \eqref{eqapp:QSL-dynamical-P}, we see that the dynamical part gives the same contribution (at this order). Thus, we conclude that MUS have a larger temporal uncertainty precisely because of the initial entanglement (last term of \eqref{eq_appC:QSL-M-dynamical}).

\section{Space--time trade-off}\label{app:trade-off}

Eq. \eqref{eq:X-variance-free-final} in the main text is the minimum spatial spread of a free, composite particle with a given internal-energy spread $\Delta_\Psi \hat{H}_c$. Achieving
this minimum requires entanglement between the clock and its c.o.m.; however, our analysis of temporal uncertainty (Eqs. \eqref{eq:QSL-generalized-static},\eqref{eq:QSL-qubit-static} in the main text) shows that the same entanglement induces decoherence in the internal-energy basis and thus worsens temporal precision . 
Specifically, in Appendices~\ref{app:MUS-states} and \ref{app:clock-precision} we analysed the spatial and temporal uncertainties for two classes of states: standard Gaussian states (Eq.~\eqref{eq:MUS-P-state}) and the MUS (Eq.~\eqref{eqapp:MUS-C-state}). We observed that, while MUS achieve the minimum spatial spread and hence are more localized in space, they exhibit a larger minimal temporal uncertainty due to the initial entanglement between the clock and its c.o.m. degrees of freedom. In the following we make this statement quantitative. Throughout this appendix, all expressions are understood up to corrections of order
$o\!\left(\big(\Delta_\Psi \hat{H}_c/mc^2\big)^{\!2}\right)$..\\

\paragraph*{Spatial uncertainty.}
Denote by $\Delta^2_\Psi \hat{x}(t)^{M}_{\min}$ and $\Delta^2_\Psi \hat{x}(t)^{G}_{\min}$
the minimal position variances \footnote{Specifically, the minimum is taken with respect to the spread in velocity $\Delta_\Psi \hat{v}$, which is a free parameter (see Eq. ~\eqref{eq:variance-X-not-minimized}.} attainable within the MUS and Gaussian families, respectively. From Eq.~\eqref{eq:X-variance-free-final} we have
\begin{equation}\label{eq_appD:X-variance-M}
  \Delta^2_\Psi \hat{x}(t)^{M}_{\min}
  \;=\;
  \frac{\hbar|t|}{\overline{m}}
  \left(1+\frac{\Delta_\Psi^2 \hat{H}_c}{m^2c^4}\right),
\end{equation}
whereas from Eq.~\eqref{eq:X-variance-P-final} we have
\begin{equation}\label{eq_appD:X-variance-G}
  \Delta^2_\Psi \hat{x}(t)^{G}_{\min}
  \;=\;
  \frac{\hbar|t|}{\overline{m}}
  \left(1+\frac{3}{2}\frac{\Delta_\Psi^2 \hat{H}_c}{m^2 c^4}\right)
  +\frac{\Delta_\Psi^2 \hat{H}_c}{m^2 c^4}\left(\frac{p_0\,t}{m}\right)^{\!2},
\end{equation}
where $\overline{m} \coloneq m+\langle \hat{H}_\mathrm{c}\rangle_\Psi / c^2$ denotes the renormalized mass. Comparing Eqs.~\eqref{eq_appD:X-variance-M} and \eqref{eq_appD:X-variance-G} shows that
standard Gaussian states exhibit a larger minimal spatial spread. More precisely,
\begin{equation}\label{eq_app:price-to-pay-x}
  \Delta^2_\Psi \hat{x}(t)^{G}_{\min}-\Delta^2_\Psi \hat{x}(t)^{M}_{\min}
  \;=\;
  \frac{\hbar|t|}{\overline{m}}\,
  \frac{\Delta_\Psi^2 \hat{H}_c}{m^2c^4}\left(
  \frac{1}{2} + \frac{p_0^2|t|}{\hbar m}\right),
\end{equation}
where the second term is the kinematical ($p_0$-dependent) contribution.\\

\paragraph*{Temporal uncertainty.}
We quantify clock precision via the refined quantum speed limit
(Eq.~\eqref{eq_app:QSL_refined}) and, for concreteness, specialise to a qubit clock
\footnote{As discussed in Appendix~\ref{app:C2-QSL}, these results extend straightforwardly to higher dimensional clocks.}.
Since we will compare with spatial spreads expressed as variances, we work with the
corresponding squared QSL time scale. Let $t^{2,M}_{\delta,\min}$ and $t^{2,G}_{\delta,\min}$
denote the minimal squared distinguishability times within the two families.
For MUS, Eq.~\eqref{eq_appC:QSL-M-dynamical} gives

\begin{gather}\label{eq_appD:QSL-M}
  t^{2,M}_{\delta,\min}
  =
  \left(\frac{\delta \hbar}{\Delta_\Psi \hat{H}_c}\right)^{\!2}
  \Bigg[
  1+\frac{\Delta^2_\Psi \hat{p} + p_0^2}{m^2c^2}\nonumber \\
  +\frac{\delta^2}{3}\left(\frac{2\Delta^2_\Psi \hat{p}}{m^2c^2}\right)^{\!2}
  \left(\frac{1}{2}+\frac{p_0^2}{\Delta^2_\Psi \hat{p}}\right)
  +\frac{\Delta_\Psi^2 \hat{H}_c}{m^2c^4}\left(\frac{1}{2}+\frac{p_0^2}{\Delta^2_\Psi \hat{p}}\right)
  \Bigg].
\end{gather}
The last term in Eq.~\eqref{eq_appD:QSL-M} is precisely
the contribution due to the initial clock--c.o.m.\ entanglement. For standard Gaussian states, which are initially separable, Eq. \eqref{eqapp:QSL-dynamical-P} gives
\begin{gather}\label{eq_appD:QSL-G}
  t^{2,G}_{\delta,\min}
  =
  \left(\frac{\delta \hbar}{\Delta_\Psi \hat{H}_c}\right)^{\!2}
  \Bigg[
  1+\frac{\Delta^2_\Psi \hat{p} + p_0^2}{m^2c^2}
  \nonumber \\ +\frac{\delta^2}{3}\left(\frac{2\Delta^2_\Psi \hat{p}}{m^2c^2}\right)^{\!2}
  \left(\frac{1}{2}+\frac{p_0^2}{\Delta^2_\Psi \hat{p}}\right)
  \Bigg].
\end{gather}
Therefore MUS display a larger minimal temporal uncertainty, entirely due to the initial entanglement:
\begin{equation}\label{eq_app:price-to-pay-t}
  t^{2,M}_{\delta,\min}-t^{2,G}_{\delta,\min}
  \;=\;
  \left(\frac{\delta \hbar}{\Delta_\Psi \hat{H}_c}\right)^{\!2}
  \frac{\Delta_\Psi^2 \hat{H}_c}{m^2c^4}
  \left(\frac{1}{2}+\frac{p_0^2}{\Delta^2_\Psi \hat{p}}\right).
\end{equation}

Eqs.~\eqref{eq_app:price-to-pay-x} and \eqref{eq_app:price-to-pay-t} show that the gain in
spatial localisation for MUS comes with an increase in temporal uncertainty. Both differences
split into (i) a ``rest-frame'' contribution (the $1/2$ term), which survives for $p_0=0$, and
(ii) a kinematical ($p_0$-dependent) contribution. We now summarise these two effects as
space--time trade-offs.\\

\paragraph*{Rest-frame trade-off.}
Setting $p_0=0$, we introduce the dimensionless quantities
\begin{equation}
  \widetilde{\Delta^2 t}
  :=
  \left(\frac{\Delta_\Psi \hat{H}_c}{\delta\hbar}\right)^{\!2} t^{2}_{\delta,\min},
  \qquad
  \widetilde{\Delta^2 x}
  :=
  \left(\frac{\overline m}{\hbar|t|}\right)\Delta^2_\Psi \hat{x}(t)_{\min}.
\end{equation}
Then Eqs.~\eqref{eq_app:price-to-pay-x} and \eqref{eq_app:price-to-pay-t} imply
\begin{equation}
  \widetilde{\Delta^2 x}^{\,G}-\widetilde{\Delta^2 x}^{\,M}
  \;=\;
  \widetilde{\Delta^2 t}^{\,M}-\widetilde{\Delta^2 t}^{\,G}
  \;=\;
  \frac{1}{2}\frac{\Delta_\Psi^2 \hat{H}_c}{m^2c^4},
\end{equation}
and therefore the simple trade-off
\begin{equation}\label{eq_appD:trade-off-rest-frame}
  \widetilde{\Delta^2 t}^{\,G}+\widetilde{\Delta^2 x}^{\,G}
  \;=\;
  \widetilde{\Delta^2 t}^{\,M}+\widetilde{\Delta^2 x}^{\,M}.
\end{equation}
Equation~\eqref{eq_appD:trade-off-rest-frame} shows that, once expressed in terms of the dimensionless quantities above, the improvement in spatial localisation provided by MUS is exactly compensated by a loss in temporal precision, and vice versa.\\

\paragraph*{Kinematical trade-off.}
We now focus on the terms that depend on the c.o.m.\ momentum $p_0$.
For standard Gaussian states the kinematical contribution appears in the spatial
spread; from Eq.~\eqref{eq_appD:X-variance-G} we can isolate
\begin{equation}
  \Delta^2_\Psi \hat{x}(t)^{G}
  \,\ge\,
  \frac{\Delta_\Psi^2 \hat{H}_c}{m^2 c^4}\left(\frac{p_0\,t}{m}\right)^{\!2}.
\end{equation}
On the other hand, the QSL implies $t_{\delta} \;\ge\; \delta \hbar/\Delta_\Psi \hat{H}_c$. Combining these two bounds yields
\begin{equation}\label{eq_appD:kinematical-trade-off}
  \Delta_\Psi \hat{x}(t)\, t_\delta
  \,\geq\,
  \frac{\hbar\,\delta}{mc}\,\frac{|p_0\, t|}{mc},
\end{equation}

For MUS states the same relation~\eqref{eq_appD:kinematical-trade-off} holds, but the
kinematical contribution is instead carried by the temporal uncertainty. From  Eq.~\eqref{eq_appD:QSL-M}) we can isolate the inequality
\begin{gather}
  t^{2,M}_{\delta} \geq  \left(\frac{\delta \hbar}{mc^2}\right)^{\!2}
 \left(\frac{1}{2}+\frac{p_0^2}{\Delta^2_\Psi \hat{p}}\right)
\end{gather}
while the spatial spread is always set by the canonical bound $\Delta^2_\Psi \hat{x}(t)\geq\hbar^2/(4\Delta^2_\Psi \hat{p})$. Combining the two, we recover \eqref{eq_appD:kinematical-trade-off}.
Thus, depending on the chosen class of states, the kinematical contribution originating
from special-relativistic time dilation manifests either in the spatial spread or in the
temporal uncertainty (or in both), which is precisely what occurs for standard Gaussian
and MUS states, respectively.

For an ideal clock, where proper time is represented by a self-adjoint operator $\hat{\tau}_c$
canonically conjugate to the internal Hamiltonian, $[\hat{\tau}_c,\hat{H}_c]=i\hbar$
(see also Sec.~\ref{sec:internal-perspective}), this uncertainty relation becomes particularly transparent. Using the Heisenberg position operator
$\hat{x}(t)=\hat{x}+\hat{p}t/\hat{m}$ one finds
\begin{equation}
  \big[\hat{x}(t), \hat{\tau}_c\big]
  = i\hbar\,\frac{\hat{p}\,t}{\hat{m}^2 c^2},
\end{equation}
and hence an uncertainty relation completely analogous to Eq.~\eqref{eq_appD:kinematical-trade-off}:
\begin{equation}
  \Delta_\Psi \hat{x}(t)\, \Delta_\Psi\hat{\tau}_c
  \,\geq\, \frac{\hbar}{mc}\,\frac{|p_0 \,t|}{2mc}.
\end{equation}

\section{Covariant POVM}\label{app:covariant-POVM}

\noindent In this appendix, we construct a POVM for the STQRF $\mathcal H_{rc} = \mathcal{H}_r\otimes \mathcal{H}_c$ which is covariant under space- \emph{and} time-translations. The generators of (1$D$) space and time translations are respectively, the total momentum $\hat{p}_r$ and the low-energy Hamiltonian 

\begin{gather}\label{eqapp:POVM-low-energy-hamiltonian}
\hat{H}_{rc}=\frac{\hat{p}_r^2}{2\hat{m}}+\hat{m}c^2,\qquad \hat{m}:=m_r+\frac{\hat{H}_c}{c^2},
\end{gather}
whose spectrum reads

\begin{equation}\label{eqapp:POVM-low-energy-hamiltonian-spectrum}
E(p,\epsilon):=\bra{p}_r \, \bra{\epsilon}_c\, \hat{H}_{rc}\,\ket{p}_r\,\ket{\epsilon}_c
=\frac{p^2}{2m(\epsilon)}+m(\epsilon)c^2
\end{equation}
where $m(\epsilon)=m+\epsilon/c^2$. For $(x,t)\in\mathbb R^2$ we define:

\begin{equation}\label{eqapp:space-time-translations-STQRF}
\hat{U}_{rc}(x,t):=\exp\Big(-\frac{i}{\hbar}\big(x\hat{p}_r+t\hat{H}_{rc}\big)\Big).
\end{equation}
A covariant POVM can be constructed starting from an arbitrary POVM element $\hat{E}_0$, the ``seed'', and considering its orbit under the group. In our case:

\begin{gather}\label{appeq:covariant-POVM-seed}
   \hat{E}(x,t) = \hat{U}^\dagger_{rc}(x,t) \, \hat{E}_0\, \hat{U}_{rc}(x,t)
\end{gather}
This ensures that the POVM is covariant under space-time translations \eqref{eqapp:space-time-translations-STQRF}. The (positive) seed $\hat{E}_0$ must be chosen such that the resulting POVM is well-defined, in particular normalized:

\begin{gather}\label{eqapp:normalization-condition-POVM}
    \int_{\mathbb{R}^2} dx\, dt \,\hat{E}(x,t) = \mathbb{I}_{r}\otimes\mathbb{I}_{c}
\end{gather}
In the following, we use the Fourier conventions $\braket{x}{p}=(2\pi\hbar)^{-1/2}e^{+\frac{i}{\hbar}px}$ and
$\braket{t}{\epsilon}=(2\pi\hbar)^{-1/2}e^{+\frac{i}{\hbar}\epsilon t}$. We will show that a well defined covariant POVM can be constructed out of the following seed state:

\begin{equation}\label{POVM-seed-all-orders}
\hat{E}_0\;=\;\hat{\Delta}^{1/2}\,\Big(\ketbra{x_0}{x_0}_{r}\otimes\ketbra{\tau_0}{\tau_0}_{c}\Big)\,\hat{\Delta}^{1/2},
\end{equation}
where $\hat{\Delta}$ is the time dilation (or redshift) operator 

\begin{gather}
    \hat{\Delta}\;:=\;\frac{\partial \hat{H}_{rc}}{\partial \hat{H}_c} = 1-\frac{\hat{p}_r^2}{2\hat{m}^2c^2}
\end{gather}
with spectrum

\begin{gather}
    \Delta(p,\epsilon) = \frac{\partial E(p,\epsilon)}{\partial \epsilon} = 1-\frac{p^2}{2m(\epsilon)^2c^2}
\end{gather}

\subsection{Non-relativistic limit}

In the non-relativistic limit $c\to \infty$ internal and external d.o.f. decouple and the Hamiltonian reads 

\begin{gather}\label{non-relativistic-hamiltonian}
    \hat{H}_{n.r.}=\frac{\hat{p}_r^2}{2m_r}+\hat{H}_{c}
\end{gather}
In this case, a covariant POVM can be constructed using uncorrelated 1D projectors on position and time space, i.e.

\begin{gather}\label{POVM-non-relativistic-seed}
   \hat{E}_0 = \ketbra{x_0}{x_0}_r \otimes \ketbra{\tau_0}{\tau_0}_c
\end{gather}
One can check that the normalization condition is satisfied. Physically, this POVM measures the time of the clock and the position of the kinematical d.o.f. The parameters $x_0, \tau_0$ are arbitrary: they represent the condition that the (proper) time according to the clock is $\tau_0$ and the c.o.m.'s position is $x_0$.
With the low-energy Hamiltonian of Eq. \eqref{eqapp:POVM-low-energy-hamiltonian}, the normalization condition is no longer satisfied. The seed element of Eq. \ref{POVM-non-relativistic-seed} must be modified to take into account relativistic corrections.

\begin{proof}
     We start by expanding both projectors in position and energy space, i.e. $\ket{x} = (2\pi\hbar)^{-1/2}\int_{\mathbb{R}} dp\, e^{-\frac{i}{\hbar}px} \ket{p}$ and $\ket{t} = (2\pi\hbar)^{-1/2}\int_{\mathbb{R}} d\epsilon\, e^{-\frac{i}{\hbar}\epsilon t} \ket{\epsilon}$. Then, using the low-energy Hamiltonian defined in Eq. \eqref{eqapp:POVM-low-energy-hamiltonian}, POVM elements read:

    \begin{gather}
       \hat{E}(x,t) = (2\pi\hbar)^{-2} \int_{\mathbb{R}^2} dp\, dp'\, d\epsilon\, d\epsilon' \, e^{-\frac{i}{\hbar}(x+x_0)(p-p')} \, \nonumber 
        \\e^{-\frac{i}{\hbar}t(\frac{p^2}{2m(\epsilon)}-\frac{p'^2}{2m(\epsilon')})}  e^{-\frac{i}{\hbar}(t+\tau_0)(\epsilon\,-\epsilon'\,)}\ketbra{p}{p'}_{r} \otimes \ketbra{\epsilon}{\epsilon'}_{c}\nonumber
    \end{gather}

    The integration over $x$ gives $2\pi \hbar\,\delta(p-p')$:

    \begin{gather}
        \int_{\mathbb{R}} dx\, \hat{E}(x,t) = (2\pi\hbar)^{-1}\int_{\mathbb{R}} dp\int_{\mathbb{R}^2} d\epsilon\, d\epsilon'    \int_{\mathbb{R}} dt\,e^{-\frac{i}{\hbar}(t+\tau_0)(\epsilon-\epsilon')}\nonumber 
        \\  e^{-\frac{i}{\hbar}t\, \frac{p^2}{2}\big(\frac{1}{m(\epsilon)}-\frac{1}{m(\epsilon')}\big)} \,\ketbra{p}{p}_{r} \otimes \ketbra{\epsilon}{\epsilon'}_{c} \nonumber
    \end{gather}
In the non-relativistic limit, $m(\epsilon) \to m_r$ so the integration over time yields a simple delta function in energy space and, consequently, the POVM is normalized. Considering relativistic corrections, this is not the case. For instance, consider the first order expansion $m(\epsilon)^{-1} \simeq m^{-1}(1-\frac{\epsilon}{mc^2})$. The above expression becomes:

    \begin{gather}
     \int_{\mathbb{R}} dx \,  \hat{E}(x,t) = (2\pi\hbar)^{-1}\,\int_{\mathbb{R}} dp \int_{\mathbb{R}^2}d\epsilon\, d\epsilon' \, e^{-\frac{i}{\hbar}(t\Delta(p)+\tau_0)(\epsilon\,-\epsilon'\,)} \, \,\nonumber 
        \\ \ketbra{p}{p'}_{r} \otimes \ketbra{\epsilon}{\epsilon'}_{c}\nonumber 
    \end{gather}
After a change of variables $ t\Delta(p)+\tau_0 \to t$ we perform the integration over $t$, which gives $2\pi\hbar\,\delta(\epsilon-\epsilon')$, resulting in 

    \begin{gather}
        \int_{\mathbb{R}^2} dx\, dt\, \hat{E}(x,t) = \int_{\mathbb{R}} dp\,\frac{1}{\Delta(p)}\, \ketbra{p}{p}_{r} \otimes \int_{\mathbb{R}} d\epsilon \, \ketbra{\epsilon}{\epsilon}_{c} \neq \mathbb{I}_{r} \otimes \mathbb{I}_{c}\nonumber 
    \end{gather}

\end{proof}

\subsection{First-order expansion}

Consider a first-order expansion in powers of $\hat{H}_{c}/mc^2$: 

\begin{gather}\label{eqapp:POVM-Ham-first-order}
    \hat{H}_{rc} = \frac{\hat{p}_r^2}{2m_r} + \hat{H}_c\left(1-\frac{\hat{p}_r^2}{2m_r^2c^2}\right) \coloneq \frac{\hat{p}_r^2}{2m_r} + \hat{H}_c \, \Delta(\hat{p}_r)
\end{gather}
Physically, the internal degrees of freedom evolve with respect to the proper time of the clock $\tau(\hat{p}_r) = t\, \Delta(\hat{p}_r)$. Intuitively, we have to convert the sharp POVM labelled by the proper time $\tau$ into one labelled by the external time parameter $t$:

\[\int_{\mathbb{R}} d\tau\, \hat U^\dagger(\tau)\hat E_0 \hat U(\tau)
\;\to\;
\int_{\mathbb{R}} dt\, \frac{d\tau}{dt}\,\hat U^\dagger(\tau(t))\hat E_0 \hat U(\tau(t)).\]
The Jacobian $d\tau/ dt = \Delta(\hat{p}_r)$ is operator valued and does not commute with $\hat{E}_0$. In order to ensure positivity and to avoid ordering ambiguities, we define:

\begin{gather}\label{covariant-POVM-seed-first-order}
    \hat{E}_0 =  \Delta(\hat{p}_r)^{1/2} \left( \ketbra{x_0}{x_0}_{r}\, \otimes \, \ketbra{\tau_0}{\tau_0}_{c} \right)\, \Delta(\hat{p}_r)^{1/2}
\end{gather}
One can check that the normalization condition in Eq. \eqref{eqapp:normalization-condition-POVM} is satisfied:

\begin{proof}
     We follow the previous proof, but using the Hamiltonian of Eq. \eqref{eqapp:POVM-Ham-first-order}:

    \begin{gather}
       \hat{E}(x,t) = (2\pi\hbar)^{-2}\int_{\mathbb{R}^2} dp\, dp'\int_{\mathbb{R}^2} d\epsilon\, d\epsilon' \, e^{-\frac{i}{\hbar}(x+x_0)(p-p')} \, e^{-\frac{i}{\hbar}t\frac{p^2-p'^2}{2m_r}} \nonumber \\ e^{-\frac{i}{\hbar}t(\epsilon\, \Delta(p)-\epsilon'\, \Delta(p'))} \, e^{-\frac{i}{\hbar}\tau_0(\epsilon -\epsilon')}\Delta(p)^{1/2} \,\Delta(p')^{1/2} \, \nonumber
        \\ \ketbra{p}{p'}_{r} \otimes \ketbra{\epsilon}{\epsilon'}_{c}\nonumber 
    \end{gather}
The integration over $x$ gives $2\pi\hbar\,\delta(p-p')$:

    \begin{gather}
        \int_{\mathbb{R}^2} dx\, dt\, \hat{E}(x,t) = (2\pi\hbar)^{-1}\int_{\mathbb{R}} dp\,\int_{\mathbb{R}^2} d\epsilon\, d\epsilon'  \,  \, \Delta(p) \nonumber \\
        \,  \int_{\mathbb{R}} dt\,e^{-\frac{i}{\hbar}(t\Delta(p)+\tau_0) 
        (\epsilon-\epsilon') }   \ketbra{p}{p}_{r} \otimes \ketbra{\epsilon}{\epsilon'}_{c}\nonumber 
    \end{gather}
After a change of variables $t\to t\Delta(p)$ we perform the integration over $t$, which gives a $2\pi\hbar\, \delta(\epsilon-\epsilon')$, resulting in 

    \begin{gather}
       \int_{\mathbb{R}^2} dx\,dt\, \hat{E}(x,t) = \int_{\mathbb{R}} dp\, \ketbra{p}{p}_{r} \otimes \int_{\mathbb{R}} d\epsilon \, \ketbra{\epsilon}{\epsilon}_{c} = \mathbb{I}_{r} \otimes \mathbb{I}_{c}\nonumber 
    \end{gather}
\end{proof}

\subsection{Second-order expansion}

Similarly, we can consider a second order expansion 

 \begin{gather}\label{eqapp:POVM-Ham-second-order}
    \hat{H}_{rc} = \frac{\hat{p}_r^2}{2m_r} + \hat{H}_c\left(1-\frac{\hat{p}_r^2}{2m_r^2c^2}\right) + \frac{\hat{H}_c^2}{m_r^2c^4} \, \frac{\hat{p}_r^2}{2m_r}
\end{gather}
In this case, a covariant POVM can be constructed from the following positive ``seed'' element:

\begin{gather}\label{covariant-POVM-seed-second-order}
    \hat{E}_0 =  \hat{\Delta}^{\frac{1}{2}} \, \biggl( \ketbra{x_0}{x_0}_{r} \otimes \ketbra{\tau_0}{\tau_0}_{c}\biggr)  \hat{\Delta}^{\frac{1}{2}} 
\end{gather}
which has the same form of Eq. \eqref{covariant-POVM-seed-first-order}, but the time dilation factor now entails the first order relativistic corrections to the mass:

\begin{gather}\label{time_dilation_factor_second_order}
    \hat{\Delta} = 1-\frac{\hat{p}_r^2}{2m_r^2c^2}\biggl(1-2\frac{\hat{H}_c}{m_rc^2}\biggr)\equiv 1-\frac{\hat{p}_r^2}{2\hat{m}^2c^2}
\end{gather}
where last equality uses the first order expansion $\hat{m}^{-2} \simeq m^{-2}\,(1-2\hat{H}_c/m_rc^2)$.

We stress that considering second order terms in the Hamiltonian is necessary if we want to look at how the internal-energy spread influence the dynamics of the c.o.m. (see section \ref{sec:external-perspective}).

\begin{proof}
We follow the previous calculations, but using the Hamiltonian of Eq. \eqref{eqapp:POVM-Ham-second-order}. The integration over $x$ always gives a $2\pi\hbar\,\delta(p-p')$. Then, we change variables $(\epsilon,\epsilon') \to (e,E) = (\epsilon-\epsilon', \frac{\epsilon+\epsilon'}{2})$, so that $\epsilon^2-\epsilon'^2 = 2eE$. This gives:

\begin{gather}
    \int_{\mathbb{R}^2} dx\, dt\, \hat{E}(x,t) =(2\pi\hbar)^{-1} \int_{\mathbb{R}} dp\,\int_{\mathbb{R}^2} de \, dE  \, \nonumber \\
    \Big(\Delta(p,E+\frac{e}{2})\, \Delta(p,E-\frac{e}{2})\Big)^{1/2}\, \nonumber 
    \\ \int_{\mathbb{R}} dt\,e^{-\frac{i}{\hbar}\,e(t\Delta(p)+\tau_0)}  e^{-\frac{i}{\hbar}t(\frac{p^2}{2m_r}\, \frac{2eE}{m^2c^4})} \ketbra{p}{p}_{r} \otimes \ketbra{E+\frac{e}{2}}{E-\frac{e}{2}}_{c}\nonumber 
\end{gather}
Finally, we rearrange the exponent $e\,\Delta(p) - e\,\frac{p^2}{2m_r}\, \frac{2E}{m^2c^4} = e\, (1-\frac{p^2}{2m_r^2c^2}(1- \frac{2E}{m_r c^2})) = e\, \Delta(p,E)$, as defined in Eq. \eqref{time_dilation_factor_second_order}. After a change of variables $t\to t\Delta(p,E)$ we perform the integration over $t$, which gives a $2\pi\hbar\,\delta(e)$, resulting in 

    \begin{gather}
        \int_{\mathbb{R}^2} dx\, dt\, \hat{E}(x,t) = \int_{\mathbb{R}} dp\, \ketbra{p}{p}_{r} \otimes \int_{\mathbb{R}} dE \, \ketbra{E}{E}_{c} = \mathbb{I}_{r} \otimes \mathbb{I}_{c}\nonumber 
    \end{gather}

\end{proof}

\subsection{All–orders}\label{subsec:all-orders-normalization}

From the Hamiltonian in Eq. \eqref{eqapp:POVM-low-energy-hamiltonian}, we define the time dilation operator:

\begin{gather}
   \hat{ \Delta}\;:=\;\frac{\partial \hat{H}_{rc}}{\partial \hat{H}_c} = 1-\frac{\hat{p}_r^2}{2\hat{m}^2c^2}
\end{gather}

which we assume to be positive in the weak–relativistic regime, i.e where $p \ll m(\epsilon)\,c$. Then, defining the ``seed'' element as

\begin{equation}\label{eq:seed-exact}
\hat{E}_0\;=\;\hat{\Delta}^{1/2}\,\Big(\ketbra{x_0}{x_0}_{r}\otimes\ketbra{\tau_0}{\tau_0}_{c}\Big)\,\hat{\Delta}^{1/2},
\end{equation}
the normalization condition of Eq. \eqref{eqapp:normalization-condition-POVM} is satisfied at all orders in $\hat{H}_c/mc^2$.

\begin{proof}

From the spectrum defined in Eq. \eqref{eqapp:POVM-low-energy-hamiltonian-spectrum} we define the phase difference
\(
\Phi_p(\epsilon,\epsilon'):=E(p,\epsilon)-E(p,\epsilon').
\). The POVM elements read:

\begin{gather}
E(x,t)
=(2\pi\hbar)^{-2}\,\!\!\int_{\mathbb{R}^2}\! dp\,dp'\int_{\mathbb{R}^2}d\epsilon\,d\epsilon'\;
e^{-\frac{i}{\hbar}(x+x_0)(p-p')}
e^{-\frac{i}{\hbar}t\,\Phi_p(\epsilon,\epsilon')} \nonumber \\\;e^{-\frac{i}{\hbar}\tau_0(\epsilon -\epsilon')}
\Delta(p,\epsilon)^{1/2}\Delta(p',\epsilon')^{1/2}\;
\ketbra{p}{p'}_r\otimes\ketbra{\epsilon}{\epsilon'}_c
\end{gather}
Integrating over $x$ gives $(2\pi\hbar)\delta(p-p')$. After integrating in $p$, we make the same change of variables of previous proof, namely $\epsilon=E+\frac{e}{2},\ \ \epsilon'=E-\frac{e}{2}$. 
By the fundamental theorem of calculus,
\begin{equation}\label{eqapp:phase-difference}
\Phi_p(e,E)
=\int_{E-\frac{e}{2}}^{E+\frac{e}{2}}\frac{\partial E(p,\eta)}{\partial \eta}\,d\eta
=\int_{E-\frac{e}{2}}^{E+\frac{e}{2}}\Delta(p,\eta)\,d\eta
=\; e\;\overline{\Delta},
\end{equation}
where
\(
\overline{\Delta}:=\frac{1}{e}\int_{E-\frac{e}{2}}^{E+\frac{e}{2}}\Delta(p,\eta)\,d\eta
\) and, by continuity, $\overline{\Delta}(e=0)=\Delta(p,E)$. Using Eq. \eqref{eqapp:phase-difference}, the integration over time yields:

\begin{equation}\label{eq:t-int}
\int_{\mathbb{R}} dt\;e^{-\frac{i}{\hbar}t\,e\,\overline{\Delta}}
=2\pi\hbar\;\delta\!\big(e\,\overline{\Delta}\big)
=\frac{2\pi\hbar}{|\overline{\Delta}|}\,\delta(e).
\end{equation}
We recall that we are assuming $\hat{\Delta}>0$, which is physically justified in the low-energy regime we are working. Finally, the $\delta(e)$ sets $e=0$, hence the prefactor becomes $\Delta(p,E)$, while $|\overline{\Delta}(e=0)|^{-1}=|\Delta(p,E)|^{-1}$, so they cancel out: 

\begin{equation}
\int_{\mathbb{R}^2} dx\,dt\,E(x,t)
=\int_{\mathbb{R}^2} dp\,dE \; \ketbra{p}{p}_{r}\otimes\ketbra{E}{E}_{c}
=\mathbb I_{r}\otimes\mathbb I_{c},
\end{equation}

\end{proof}

\section{Relational position operator}\label{app:relative-position-operator}

\noindent We consider the relational operator conditioned on the STQRF's readout \((x_0,\tau_0)\), defined in Eq. \ref{eq:relational-position-operator-general} in the main text

\begin{equation}\label{eq:relpos-def-external-exp-tau0}
  \hat{\widetilde{x}}_{(x_0,\tau_0)}
=\int_{\mathbb{R}^2} dx\,dt\; \hat{E}(x,t)\,\otimes\, \left(\hat{x}_s - \frac{\hat{p}_s}{m_s}t -x\right),
\end{equation}
where the readout is encoded in the POVM elements
\begin{gather}
    \hat{E}(x,t)=\hat{U}^\dagger_{rc}(x,t)\, \Big[\hat{\Delta}^{1/2}\Bigl(\ketbra{x_0}{x_0}_{r}\otimes\ketbra{\tau_0}{\tau_0}_{c}\Bigr)\hat{\Delta}^{1/2}\Big]\,\hat{U}_{rc}(x,t).
\end{gather}
and the action of space-time translations on the STQRF reads

\begin{gather}
    \hat{U}_{\mathrm{rc}}(x,t) = e^{-\tfrac{i}{\hbar}\big(x\,\hat{p}_{\mathrm{r}}+t\,\hat{H}_{\mathrm{rc}}\big)}.
\end{gather}

In the following we first perform the spatial twirl and then the temporal one; the order is not relevant since space and time translations commute.
\\

\paragraph*{Spatial twirl (integration over $x$).}
We expand
\[
\ketbra{x_0}{x_0}_{r}
=\int_{\mathbb{R}^2}\!\frac{dp\,dp'}{2\pi\hbar}\,e^{-\frac{i}{\hbar}x_0(p-p')}\,\ketbra{p}{p'}_{r},
\]
and define the operator $\hat{\Delta}(p):=\bra{p}\,\hat{\Delta}\,\ket{p}_r$ acting on $\mathcal{H}_c$. The integrand of the $x$–twirl (after shifting the integration $x \to x+x_0$) is

\begin{gather}
    \int_{\mathbb{R}^2}\!\frac{dp\,dp'}{2\pi\hbar}\ e^{-\frac{i}{\hbar}x(p-p')} \, \ketbra{p}{p'}_{r} \nonumber \\
     \otimes\ \hat{\Delta}(p)^{1/2}\ketbra{\tau_0}{\tau_0}_{c}\hat{\Delta}(p')^{1/2}
\ \otimes\ \Bigl(\hat{x}_s  - \frac{\hat{p}_s}{m_s}\,t+x_0-x\Bigr).
\end{gather}
Consider two contributions separately:\\

\emph{(i) \(\hat{x}_s - \tfrac{t}{m_s}\hat{p}_s + x_0 \):} Using
\(\int_{\mathbb{R}} dx\,e^{-\frac{i}{\hbar}x(p-p')}=2\pi\hbar\,\delta(p-p')\), the integral over $x$ yields
\begin{gather}
\int_{\mathbb{R}} dp\;\ketbra{p}{p}_{r}\otimes\hat{\Delta}(p)^{1/2}\ketbra{\tau_0}{\tau_0}_{c}\hat{\Delta}(p)^{1/2}\otimes \left( \hat{x}_s - \tfrac{t}{m_s}\hat{p}_s + x_0\right).
\end{gather}

\emph{(ii) \(x\):} Here, we change variables \(p=Q+\tfrac{q}{2}\), \(p'=Q-\tfrac{q}{2}\) and use \(\int_{\mathbb{R}} dx\,x\,e^{-\frac{i}{\hbar}xq}=+\,i\,2\pi\,\hbar^2\,\delta'(q)\):
\begin{gather}
    +\,i\,\hbar\!\int_{\mathbb{R}^2}\!dQ\,dq\ \delta'(q) \nonumber \\ \ketbra{Q+\frac{q}{2}}{Q-\frac{q}{2}}_{r} 
\otimes \hat{\Delta}(Q+\frac{q}{2})^{1/2}\ketbra{\tau_0}{\tau_0}_{c}\hat{\Delta}(Q-\frac{q}{2})^{1/2}\otimes\mathbb I_s.\nonumber
\end{gather}
Integrate by parts in \(q\) and use \(\ket{Q\pm\frac{q}{2}}_{r}=e^{\pm\frac{i}{\hbar}\frac{q}{2}\hat{x}_r}\ket{Q}_{r}\):
\[
\partial_q\bigl(\ket{Q+\tfrac{q}{2}}_{r}\bra{Q-\tfrac{q}{2}}_{r}\bigr)\Big|_{q=0}
=\frac{i}{\hbar}\,\bigl\{\hat{x}_r,\ketbra{Q}{Q}_{r}\bigr\}.
\]
where $\{\hat{A},\hat{B}\} = \frac{1}{2}(\hat{A}\hat{B}+\hat{B}\hat{A})$. Using 

\[\partial_q \, \hat{\Delta}(Q\pm q/2) = \mp \frac{Q\pm q/2}{2\hat{m}^2c^2},\]

one finds an extra contribution

\begin{gather}
   i\hbar\,  \frac{\hat{p}_r}{4c^2} \otimes \left( \hat{\Delta}^{-1/2}\frac{1}{\hat{m}^2} \ketbra{\tau_0}{\tau_0} \hat{\Delta}^{1/2} - \hat{\Delta}^{1/2}\ketbra{\tau_0}{\tau_0} \hat{\Delta}^{-1/2}\frac{1}{\hat{m}^2}\right)
\end{gather}
which arises from derivatives acting on the factors $\hat{\Delta}^{1/2}$.
Ultimately, one can check that this term vanishes once we perform the $t$-twirl, so we do not consider it in the following.
Thus, the contribution (ii) reads \(\int_{\mathbb{R}} dQ\,\{\hat{x}_r,\ketbra{Q}{Q}_{r}\}\otimes \hat{\Delta}_Q^{1/2}\ketbra{\tau_0}{\tau_0}_{c}\hat{\Delta}_Q^{1/2}\otimes\mathbb I_s\).
Putting (i)–(ii) together:
\begin{gather}
  \hat{\widetilde{x}}_{(x_0,\tau_0)} = \nonumber \\
\int_{\mathbb{R}} dp\;\ketbra{p}{p}_{r}\otimes\hat{\Delta}(p)^{1/2}\ketbra{\tau_0}{\tau_0}_{c}\hat{\Delta}(p)^{1/2}\otimes\Bigl(\hat{x}_s-t\frac{\hat{p}_s}{m_s}+x_0\Bigr)\nonumber\\
-\int_{\mathbb{R}} dp\;\bigl\{\hat{x}_r,\ketbra{p}{p}_{r}\bigr\}\otimes\hat{\Delta}(p)^{1/2}\ketbra{\tau_0}{\tau_0}_{c}\hat{\Delta}(p)^{1/2}\otimes\mathbb I_s.\label{eq:x-twirl-final}
\end{gather}

\paragraph*{Temporal twirl (integration over $t$).}
\(\hat{U}_{rc}(t)=e^{-\frac{i}{\hbar}t\hat{H}_{rc}}\) acts on both \(\ketbra{p}{p}_{r}\otimes\ketbra{\tau_0}{\tau_0}_{c}\).
Use \(\ketbra{\tau_0}{\tau_0}_c=\frac{1}{2\pi\hbar}\!\int_{\mathbb{R}^2}\!d\epsilon\,d\epsilon'\,e^{-\frac{i}{\hbar}(\epsilon-\epsilon')\tau_0}\ketbra{\epsilon}{\epsilon'}_c\),
so the integrand carries the "extra" phase \(\exp\{-\frac{i}{\hbar}(\epsilon-\epsilon')\tau_0\}\). We perform a change of variables \(\epsilon=E+\frac{e}{2}\), \(\epsilon'=E-\frac{e}{2}\) and define

\begin{equation}
\Phi(p,e,E)
=\int_{E-\frac{e}{2}}^{E+\frac{e}{2}}\frac{\partial E(p,\eta)}{\partial \eta}\,d\eta
=\int_{E-\frac{e}{2}}^{E+\frac{e}{2}}\Delta(p,\eta)\,d\eta
=\; e\;\overline{\Delta},\nonumber
\end{equation}
and from continuity, we have \(\overline\Delta(p)(e=0)=\Delta(p,E)\). Then we use
\[
\int_{\mathbb R}dt\ e^{-\frac{i}{\hbar}t\,e\,\overline{\Delta}}=2\pi\hbar\,\frac{\delta(e)}{|\overline{\Delta}|}\]
\[
\int_{\mathbb R}dt\ t\,e^{-\frac{i}{\hbar}t\,e\,\overline{\Delta}}=\, \frac{4\pi \hbar}{\Delta(p,E+\frac{e}{2})+\Delta(p,E-\frac{e}{2})} i\hbar\partial_e \left(\frac{\delta(e)}{\overline{\Delta}}\right).
\]

where the last equality follows from:

\[
\int_{\mathbb R}dt\ t\,e^{-\frac{i}{\hbar}t\,\Phi}=i\,2\pi\,\hbar^2\, \partial_\Phi \delta(\Phi)
\]
and using the chain rule $
\partial_{e}\Phi(p,e,E) = \frac{1}{2}(\Delta(p,E+\frac{e}{2})+\Delta(p,E-\frac{e}{2}))$. After integrating by parts and setting $e=0$, the factor $1/|\overline{\Delta}|$ will cancel the factor \(\Delta(p,E+\frac{e}{2})^{1/2}\Delta(p,E-\frac{e}{2})^{1/2}\to \Delta(p,E)\) at \(e=0\).\\

\emph{(A) $1^{st}$ line of Eq. \eqref{eq:x-twirl-final}.}
The \(\hat{x}_s\) term becomes \(\mathbb I_{r}\otimes\mathbb I_{c}\otimes \hat{x}_s\).
For the \(\tfrac{t}{m_s}\hat{p}_s\) term, integrate by parts in \(e\) and use
\(\ket{E\pm \tfrac{e}{2}}_{c}=e^{\pm\frac{i}{\hbar}\frac{e}{2}\hat{\tau}_c}\ket{E}_{c}\):
\[
\partial_e\Bigl(\ket{E+\tfrac{e}{2}}_{c}\bra{E-\tfrac{e}{2}}_{c}\Bigr)\Big|_{e=0}
=\frac{i}{\hbar}\bigl\{\hat{\tau}_c,\ketbra{E}{E}_{c}\bigr\}.
\]
The derivative on the extra phase \(e^{-\frac{i}{\hbar}e\,\tau_0}\) produces the replacement $\{\hat{\tau}_c,\cdot\}\ \longrightarrow\ \{\hat{\tau}_c-\tau_0,\cdot\}$, while the derivatives acting on the factor $2/\Delta(p,E+\frac{e}{2})+\Delta(p,E-\frac{e}{2})$ cancel out once we set $e=0$, since
\begin{gather}
    \partial_e \Delta(p,E\pm\frac{e}{2})\Big|_{e=0} = \pm  \frac{p^2}{2m(E)^2c^2} \frac{1}{2mc^2} \nonumber
\end{gather}
Hence
\begin{equation}\label{eq:block-sys}
\int_{\mathbb R}dt\ \text{($1^{st}$ line of Eq. \eqref{eq:x-twirl-final})}
=\ \mathbb I_{r}\otimes\mathbb I_{c}\otimes \hat{x}_s
\ +\ \mathbb I_{r}\otimes \bigl\{\hat{\tau}_c-\tau_0,\hat{\Delta}^{-1}\bigr\}\otimes \frac{\hat{p}_s}{m_s}.
\end{equation}

\emph{(B) $2^{nd}$ line of Eq. \eqref{eq:x-twirl-final}.}
The Heisenberg evolution under \(\hat{H}_{rc}\) gives

\begin{gather}
    e^{+\frac{i}{\hbar}t\hat{H}_{rc}}\,\Big(\{\hat{x}_r,\ketbra{p}{p}_{r}\} \otimes \ketbra{\epsilon}{\epsilon'}_c\Big)\,e^{-\frac{i}{\hbar}t\hat{H}_{rc}} \nonumber\\
=\Bigg(\{\hat{x}_r,\ketbra{p}{p}_{r}\}+t\,\ketbra{p}{p}_{r}\Bigl(\tfrac{\hat{p}_r}{m(\epsilon)}+\tfrac{\hat{p}_r}{m(\epsilon')}\Bigr)\Bigg)\otimes \ketbra{\epsilon}{\epsilon'}_c.\nonumber
\end{gather}
Proceeding as in (A), we obtain

\begin{gather}
\int_{\mathbb R}dt\ \text{($2^{nd}$ line of \eqref{eq:x-twirl-final})}
=\nonumber \\ \{\hat{x}_r,\mathbb I_{r}\}\otimes\mathbb I_{c}
\ +\ \mathbb I_{r}\otimes \bigl\{\hat{\tau}_c-\tau_0,(\hat{m}\,\hat{\Delta})^{-1}\bigr\}\otimes \hat{p}_r.\label{eq:block-rod}
\end{gather}

\paragraph*{Final expression.}
Subtracting Eq. \eqref{eq:block-rod} (due to the minus sign in \eqref{eq:x-twirl-final}) to Eq. \eqref{eq:block-sys} yields:

\begin{gather}
     \hat{\widetilde{x}}_{(x_0,\tau_0)}
= \hat{x}_s
+ \frac{\hat{p}_s}{m_s}\,\Bigl\{\hat{\tau}_c-\tau_0,\;\hat{\Delta}^{-1}\Bigr\}\nonumber \\
\;-\;
\Bigl(
\hat{x}_r
+ \hat{p}_{\mathrm{r}}\,
\Bigl\{\hat{\tau}_c-\tau_0,\;(\hat{m}\,\hat{\Delta})^{-1}\Bigr\}
- x_0
\Bigr),
\end{gather}
which is Eq. \eqref{eq:relational-position-rel} in the main text.

\section{Relational position uncertainty}\label{app:relational-position-spread}
Here we compute the variance of the relational position operator 

\begin{gather}
     \hat{\widetilde{x}}_{(x_0,\tau_0)}
= \hat{x}_s
+ \frac{\hat{p}_s}{m_s}\,\Bigl\{\hat{\tau}_c-\tau_0,\;\hat{\Delta}^{-1}\Bigr\}\nonumber \\
\;-\;
\Bigl(
\hat{x}_r
+ \hat{p}_{\mathrm{r}}\,
\Bigl\{\hat{\tau}_c-\tau_0,\;(\hat{m}\,\hat{\Delta})^{-1}\Bigr\}
- x_0
\Bigr),
\label{eqapp:spread--relative-position-operator}
\end{gather}
defined as

\begin{gather}\label{appeq:def-variance-Dirac-position}
    \Delta_\Psi^2   \hat{\widetilde{x}}_{(x_0,\tau_0)} 
    = \langle   \hat{\widetilde{x}}_{(x_0,\tau_0)}^2 \rangle_\Psi 
      - \langle   \hat{\widetilde{x}}_{(x_0,\tau_0)} \rangle_\Psi^2,
\end{gather}

\subsection{Non-relativistic limit}

In the non-relativistic limit, $\Delta \!\to\! \mathbb I_c \otimes \mathbb I_r$ and $\hat{m} \!\to\! m_r\,\mathbb I_c$, so that Eq. \ref{eqapp:spread--relative-position-operator} reduces to

\begin{gather}
  \hat{\widetilde{x}}^{\,\,\text{n.r.}}_{(x_0,\tau_0)} 
= \hat{x}_s + \frac{\hat{p}_s}{m_s}\,(\hat{\tau}_c-\tau_0)
\;-\;
\Bigl(\hat{x}_r + \frac{\hat{p}_{\mathrm{r}}}{m_r}\,(\hat{\tau}_c-\tau_0) - x_0\Bigr),
 \label{eqapp:spread--relative-position-operator-non-rel}
\end{gather}
All operators belonging to different subsystems commute and
the total state factorizes, $\ket{\Psi}=\ket{\psi}_{r}\otimes \ket{\phi}_c\otimes\!\ket{\chi}_s$,
so that expectation values factorize accordingly. Squaring Eq. \eqref{eqapp:spread--relative-position-operator-non-rel}
\begin{gather}
    \bigl(  \hat{\widetilde{x}}^{\,\,\text{n.r.}}_{(x_0,\tau_0)} \bigr)^2
    = (\hat{x}_s - \hat{x}_r +x_0)^2\nonumber \\
    + \!\left(\tfrac{\hat{p}_s}{m_s} - \tfrac{\hat{p}_r}{m_r}
      \right)^2 \,(\!\hat{\tau}_c-\tau_0\!)^2  \nonumber\\
      +\Bigl[
        \{\hat{x}_s,\tfrac{\hat{p}_s}{m_s}\}
        + \{\hat{x}_r,\tfrac{\hat{p}_r}{m_r}\}
      \Bigr](\!\hat{\tau}_c-\tau_0\!),
\end{gather}
and using $\langle \hat{A}^2\rangle_\Psi=\Delta_\Psi^2\hat{A}+\langle \hat{A}\rangle_\Psi^2$ and
$\mathrm{Cov}_\Psi(\hat{A},\hat{B})=\tfrac{1}{2}\langle \hat{A}\hat{B}+\hat{B}\hat{A}\rangle_\Psi-\langle \hat{A}\rangle_\Psi\langle \hat{B}\rangle_\Psi$, we can directly compute the variance (Eq. \ref{appeq:def-variance-Dirac-position}), which reads
\begin{gather}
\Delta_\Psi^2  \hat{\widetilde{x}}^{\,\,\text{n.r.}}_{(x_0,\tau_0)}
      =\Delta_\Psi^2 \hat{x}_s +\Delta_\Psi^2 \hat{x}_r \nonumber \\
         + \Bigl(\tfrac{\Delta_\Psi^2 \hat{p}_s}{m_s^2}
                 + \tfrac{\Delta_\Psi^2 \hat{p}_r}{m_r^2}\Bigr)
           \Bigl(\Delta_\Psi^2\hat{\tau}_c
                 + \langle\hat{\tau}_c-\tau_0\rangle_\Psi^2\Bigr)
          \nonumber \\
        +2\,\langle\hat{\tau}_c-\tau_0\rangle_\Psi
          \Bigl[\mathrm{Cov}_\Psi\!\left(\hat{x}_s,\tfrac{\hat{p}_s}{m_s}\right)
                + \mathrm{Cov}_\Psi\!\left(\hat{x}_r,\tfrac{\hat{p}_r}{m_r}\right)\Bigr]
\nonumber \\
        + \Bigl(\tfrac{\langle \hat{p}_s\rangle_\Psi}{m_s}
                -\tfrac{\langle \hat{p}_r\rangle_\Psi}{m_r}\Bigr)^{\!2}
           \Delta_\Psi^2\hat{\tau}_c .
\label{appeq:spread-relational-position-non-rel-full}
\end{gather}
Eq. ~\eqref{appeq:spread-relational-position-non-rel-full}
shows that the spread of the relational position is the sum of the
individual spreads of the system and of the rod, including the correlation terms
$\mathrm{Cov}_\Psi(X_i,P_i)\,\langle\hat{\tau}_c-\tau_0\rangle_\Psi$, together with a ``drift'' contribution (last line) associated with the uncertainty in the mean
relative motion. The latter represents an additional contribution of the
clock’s time uncertainty to the variance of the measured relative distance.

Now, some important observations that we will use later as well:

\begin{enumerate}[(i)]

    \item To isolate the contribution arising solely from the STQRF, we can safely neglect the uncertainty due to the system and also ignore that due to the average relative motion, as these only increase the total uncertainty. This is equivalent to taking, from the outset, the classical limit for the system $m_s \to \infty$ and starting from the operator
 \begin{gather}
   \lim_{m_s\to \infty}  \hat{\widetilde{x}}^{\,\,\text{n.r.}}_{(x_0,\tau_0)}\;=\; \hat{x}_s - \Bigl[\hat{x}_r+\tfrac{\hat{p}_r}{m_r}\bigl(\hat{\tau}_c-\tau_0\bigr)\Bigr],
 \end{gather}
 
    and then set $\langle \hat{p}_r \rangle_\Psi = 0$.
    
 \item We already discussed the role of the correlation terms in appendix~\ref{app:contractive-states}. In this paper, we assume that the states are symmetric in phase space, i.e.\ $\mathrm{Cov}_\Psi(\hat{x},\hat{p})=0$, so that the third line of Eq. \eqref{appeq:spread-relational-position-non-rel-full} vanishes identically.

 \item Physically, the average time $\langle\hat{\tau}_c\rangle_\Psi$ corresponds to the time at which we prepare the clock's state by means of an external reference frame. Thus, in a fully relational framework, it is meaningless and we conventionally set it to $\langle\hat{\tau}_c\rangle_\Psi=0$. In general, choosing a non-zero average would amount only to a redefinition of $\tau_0$, and therefore does not affect our analysis.

\end{enumerate}

Therefore, we focus on the following inequality:
\begin{gather}
   \Delta_\Psi^2 \hat{\widetilde{x}}^{\,\,\text{n.r.}}_{(x_0,\tau_0)}
      \,\geq\,\Delta_\Psi^2 \hat{x}_r
        +\frac{\Delta_\Psi^2 \hat{p}_r}{m_r^2}
          \bigl(\Delta_\Psi^2\hat{\tau}_c
                 + \tau_0^2\bigr),
\label{appeq:spread-relational-position-non-rel}
\end{gather}
which is Eq.~\eqref{eq:spread-relational-position-non-rel-rod} of the main text. The goal of the next subsection is to find a similar inequality for the full operator defined in Eq. \eqref{eqapp:spread--relative-position-operator}.

\subsection{Relativistic corrections}

The main differences with respect to the non-relativistic case are that
(i) the total mass $\hat{m}$ is now an operator that does not commute with the time
observable $\hat{\tau}_c$, and (ii) the STQRF states can be entangled through this mass operator. Explicitly,
\begin{gather}\label{appeq:QRF-state-general}
     \ket{\Psi}_{rc} =  \int_{\mathbb{R}^2} d\epsilon\,dp \;
   \psi_{v_0}(p,\epsilon)\,\phi_{\epsilon_0}(\epsilon)\,
   \ket{p}_r \otimes \ket{\epsilon}_c \nonumber \\
   = \int_{\mathbb{R}} d\epsilon \;
   \phi_{\epsilon_0}(\epsilon) \, \ket{\psi(\epsilon)}_r \otimes \ket{\epsilon}_c ,
\end{gather}
so that expectation values of operators acting on $\mathcal{H}_{rc}$ do not, in general, factorize. As a result, isolating separate contributions from the
various observables to the total spread is not straightforward.

We now proceed by making use of the three observations discussed above.

\begin{enumerate}[(i)]

\item The system $S$ is uncorrelated with the STQRF and therefore can only add
quantum uncertainty:
\[
  \Delta_\Psi^2  \hat{\widetilde{x}}_{(x_0,\tau_0)}  \geq
  \lim_{m_s\to \infty}\Delta_\Psi^2  \hat{\widetilde{x}}_{(x_0,\tau_0)} .
\]
Since we are interested solely in the contribution arising from the STQRF, we
directly take the classical limit $m_s \to \infty$, starting from
\begin{gather}\label{eqapp:spread--relative-position-operator-approx}
  \lim_{m_s\to \infty}  \hat{\widetilde{x}}_{(x_0,\tau_0)}
= \hat{x}_s - \hat{x}_r - \hat{p}_r\,\Bigl(\{\hat{\tau}_c-\tau_0,(\hat{\Delta} \hat{m})^{-1}\}\Bigr),
\end{gather}
in which case we find
\begin{gather}
        \Delta_\Psi^2  \hat{\widetilde{x}}_{(x_0,\tau_0)} 
        =\Delta_\Psi^2 \hat{x}_s + \Delta_\Psi^2 \hat{x}_r 
        + \bigl\langle  \hat{p}_r^{2}\, \hat{T}_{\hat{m},\hat{\Delta}}^{2}\bigr\rangle_\Psi \nonumber \\
        + 2\,\mathrm{Cov}_\Psi\!\bigl(\hat{x}_r, \hat{p}_r \hat{T}_{\hat{m},\hat{\Delta}}\bigr),
\label{appeq:spread-relational-position-full}
\end{gather}
where we define
$\hat{T}_{\hat{m},\hat{\Delta}} \coloneq \{\hat{\tau}_c-\tau_0,(\hat{\Delta} \hat{m})^{-1}\}$.

\item Because relativistic corrections enter only through the internal-energy–dependent mass
$\hat m(\hat H_c)$ and through the time dilation factor $\hat{\Delta}(\hat{p}_r^2)$, which depend on $\hat{p}_r$
only via even powers, they do not generate new \emph{position–momentum} correlations. It is therefore natural to retain the assumption of phase-space symmetry for the c.o.m., so that the second line in Eq.~\eqref{appeq:spread-relational-position-full} vanishes.

\item Similarly, the average time $\langle \hat{\tau}_c\rangle_\Psi=0$ is not
modified by relativistic corrections. In particular, the linear term in
$\hat{\tau}_c$ contained in $\hat{T}_{\hat{m},\hat{\Delta}}^2$ vanishes. Explicitly, considering states in the form of Eq. \eqref{appeq:QRF-state-general}, this term reads
\begin{gather}
     -2\tau_0\, \Bigl \langle \, \hat{p}_r^2 
     \Bigl\{ (\hat{\Delta} \hat{m})^{-1}, \bigl\{\hat{\tau}_c, (\hat{\Delta} \hat{m})^{-1}\bigr\} \Bigr\}\Bigr \rangle_\Psi \nonumber\\  
     = -\frac{\tau_0}{2}\int_{\mathbb{R}^2}d\epsilon d\epsilon'\;  
     \phi_{\epsilon_0}(\epsilon)\phi^*_{\epsilon_0}(\epsilon')\nonumber \\
     \bra{\psi(\epsilon)}\!\left(\frac{1}{\hat{\Delta}(\epsilon)m(\epsilon)}+\frac{1}{\hat{\Delta}(\epsilon')m(\epsilon')}\right)^{\!2}\!\hat{p}_r^2 \ket{\psi(\epsilon')}_r 
     \bra{\epsilon}\hat{\tau}_c\ket{\epsilon'},
\end{gather}
Using $\bra{\epsilon}\hat{\tau}_c\ket{\epsilon'} = -i\hbar \partial_e\delta(e)$
with $e=\epsilon-\epsilon'$, one can check that only derivatives acting on the
imaginary part of $\phi_{\epsilon_0}(\epsilon)$ contribute. This imaginary part
encodes $\langle \hat{\tau}_c\rangle_\Psi$, which vanishes by construction.

\end{enumerate}

With these points, Eq.~\eqref{appeq:spread-relational-position-full} reduces to
\begin{gather}
    \Delta_\Psi^2  \hat{\widetilde{x}}_{(x_0,\tau_0)}  \ge
     \Delta_\Psi^2 \hat{x}_r 
     + \Bigl\langle \frac{\hat{p}_r^2}{(\hat{\Delta}\hat{m})^2}
       \Bigr\rangle_\Psi \tau_0^2\nonumber \\
     + \Bigl\langle 
        \hat{p}_r^2\bigl(\{\hat{\tau}_c,(\hat{\Delta} \hat{m})^{-1}\}\bigr)^2
       \Bigr\rangle_\Psi.
\label{eqapp:relational-position-spread-not-expanded}
\end{gather}

\paragraph*{Low-energy expansion} 
We now consider relativistic corrections due to $\hat{\Delta}$ and $\hat{m}$ up
to $o(c^{-4})$, in order to separate the contributions from the various observables ($\hat{v}_r$, $\hat{x}_r$, $\hat{\tau}_c$, and
$\hat{H}_c$). We stress that we are limited to the low-energy regime, where the spectrum of
the internal Hamiltonian $\hat{H}_c$ lies well below $m_r c^2$ (equivalently,
we restrict to a subspace of the Hilbert space where this condition is
satisfied). This implies that $\Delta_\Psi \hat{H}_c \ll m_r c^2$ and, because
of the time–energy uncertainty principle (Eq.~\eqref{MT-bound-continuous-clock}), $\Delta_\Psi \hat{\tau}_c \gg
\hbar/m_r c^2$.

It is useful to introduce the dimensionless operators
\[
\hat{\varepsilon}_c \coloneq \frac{\hat{H}_c}{m_r c^2},
\qquad 
\hat{\rho}_r \coloneq \frac{\hat{p}_r}{m_r c}.
\]
The relativistic corrections in Eq.~\eqref{eqapp:relational-position-spread-not-expanded} enter through
\begin{gather}
    \frac{1}{(\hat{\Delta}\hat{m})^2} 
    = \frac{1}{m_r^2}
      \Bigl(1-2\hat{\varepsilon}_c+3\hat{\varepsilon}_c^2
            +\hat{\rho}_r^2(1-4\hat{\varepsilon}_c)
            +\tfrac{3}{4}\hat{\rho}_r^4\Bigr).
\end{gather}
We now consider the different contributions to Eq.~\eqref{eqapp:relational-position-spread-not-expanded} separately. \\

(i) The term proportional to $\tau_0^2$ is, up to $o(c^{-4})$,
\begin{gather}
     \Bigl\langle \frac{\hat{p}_r^2}{(\hat{\Delta}\hat{m})^2} 
       \Bigr\rangle_\Psi \tau_0^2 
     =  \Biggl( \bigl\langle \hat{v}_r^2\bigr\rangle_\Psi
     + \frac{m_r^4}{\overline{m}_r^{4}}c^2\bigl\langle \hat{\rho}_r^4\bigr\rangle_\Psi
     + \frac{3}{4}c^2\bigl\langle \hat{\rho}_r^6\bigr\rangle_\Psi\Biggr)  \tau_0^2 \nonumber\\
     \geq \bigl\langle \hat{v}_r^2\bigr\rangle_\Psi \tau_0^2,
\end{gather}
where $\overline{m}_r^{-1} = m_r^{-1}(1-\langle\hat{\varepsilon}_c\rangle_\Psi)$.\\

(ii) To expand the last term in Eq.~\eqref{eqapp:relational-position-spread-not-expanded} up to $o(c^{-4})$, we use the following identity:
\begin{gather}\label{eqapp:comm-rel-trick}
    \bigl\{\tfrac{1}{\hat{\Delta}\hat{m}}, \hat{\tau}_c \bigr\}^{2} 
    = \hat{\tau}_c\, \frac{1}{(\hat{\Delta}\hat{m})^{2}} \, \hat{\tau}_c
     - \frac{5}{4}\,\frac{\hbar^{2}}{(\hat{\Delta}\hat{m})^{4} c^{4}}.
\end{gather}
This follows from the canonical commutation relation $[\hat{\tau}_c,\hat{H}_c]=i\hbar$, which implies that, for any (operator-valued) function $f(\hat{m})$, we have $[\hat{\tau}_c,f(\hat{m})] = \tfrac{i\hbar}{c^2}f'(\hat{m})$. The negative term in Eq.~\eqref{eqapp:comm-rel-trick} is already $o(c^{-4})$ and will be treated consistently at the end. It follows that
\begin{gather}
    \Bigl\langle 
        \hat{p}_r^2\bigl(\{\hat{\tau}_c,(\hat{\Delta} \hat{m})^{-1}\}\bigr)^2
       \Bigr\rangle_\Psi
       = \Bigl\langle \hat{p}_r^2\,\hat{\tau}_c\,\frac{1}{\hat{m}^2}\,\hat{\tau}_c\Bigr\rangle_\Psi 
        - \frac{5}{4}\,\frac{\hbar^2}{m_r^4 c^4}\,\bigl\langle \hat{p}_r^2\bigr\rangle_\Psi \nonumber \\
        \hspace{12pt}
        + \frac{1}{m_r^2}\Biggl\langle \hat{p}_r^2\,\hat{\rho}_r^2 \,
        \hat{\tau}_c\Biggl(\!1-4\,\hat{\varepsilon}_c
        + \frac{3}{4}\!\hat{\rho}_r^4\,\Biggr)\hat{\tau}_c\Biggr\rangle_\Psi.
\end{gather}
The last line always gives a positive contribution. In fact, we have
\begin{gather}
     \frac{1}{m_r^2}\Biggl\langle \hat{p}_r^2\,\hat{\rho}_r^2 \,
        \hat{\tau}_c\Biggl(\!1-4\,\hat{\varepsilon}_c
        \Biggr)\hat{\tau}_c\Biggr\rangle_\Psi
     =  \Biggl\langle  \hat{\tau}_c\Biggl( \frac{\hat{v}_r^4}{c^2} \Biggr)\hat{\tau}_c\Biggr\rangle_\Psi \geq 0,
\end{gather}
where we used the first-order expansion
$\hat{v}_r = \tfrac{\hat{p}_r}{m_r}(1-\hat{H}_c/m_r c^2)$. To bound the first term, expand
\begin{gather}
   m_r^2\, \biggl\langle \hat{p}_r^2\,\hat{\tau}_c\,\frac{1}{\hat{m}^2}\,\hat{\tau}_c\biggr\rangle_\Psi 
   = \biggl\langle \hat{p}_r^2\,\hat{\tau}_c^2\biggr\rangle_\Psi 
    - 2\,\biggl\langle \hat{p}_r^2\,\hat{\tau}_c\,\hat{\varepsilon}_c\,\hat{\tau}_c\biggr\rangle_\Psi 
    + \nonumber \\
    3\,\biggl\langle \hat{p}_r^2\,\hat{\tau}_c\,\hat{\varepsilon}_c^2\,\hat{\tau}_c\biggr\rangle_\Psi.
\end{gather}
Now, define $\ket{u} = \hat{p}_r\,\hat{\tau}_c \ket{\Psi}$ and
$\ket{v} = \hat{p}_r\,\hat{\varepsilon}_c\,\hat{\tau}_c \ket{\Psi}$. 
Applying the Cauchy–Schwarz inequality,
\begin{gather}
    |\braket{u}{v}| \leq \|u\|\,\|v\|
    = \sqrt{\biggl\langle \hat{p}_r^2\,\hat{\tau}_c^2\biggr\rangle_\Psi 
      \biggl\langle \hat{p}_r^2\,\hat{\tau}_c\,\hat{\varepsilon}_c^2\,\hat{\tau}_c\biggr\rangle_\Psi }.
\end{gather}
We obtain
\begin{gather}
   m_r^2\,\biggl\langle \hat{p}_r^2\,\hat{\tau}_c\,\frac{1}{\hat{m}^2}\,\hat{\tau}_c\biggr\rangle_\Psi 
   \geq 
   \biggl\langle \hat{p}_r^2\,\hat{\tau}_c^2\biggr\rangle_\Psi 
   - \nonumber \\ 2\,\sqrt{\biggl\langle \hat{p}_r^2\,\hat{\tau}_c^2\biggr\rangle_\Psi 
      \biggl\langle \hat{p}_r^2\,\hat{\tau}_c\,\hat{\varepsilon}_c^2\,\hat{\tau}_c\biggr\rangle_\Psi }
   + 3\,\biggl\langle \hat{p}_r^2\,\hat{\tau}_c\,\hat{\varepsilon}_c^2\,\hat{\tau}_c\biggr\rangle_\Psi \\[4pt]
   = \frac{2}{3} \biggl\langle \hat{p}_r^2\,\hat{\tau}_c^2\biggr\rangle_\Psi
   + 3\left( \sqrt{\biggl\langle \hat{p}_r^2\,\hat{\tau}_c\,\hat{\varepsilon}_c^2\,\hat{\tau}_c\biggr\rangle_\Psi}
   -\frac{1}{3}\sqrt{ \biggl\langle \hat{p}_r^2\,\hat{\tau}_c^2\biggr\rangle_\Psi }\right)^2.
\end{gather}
so the following inequality holds:
\begin{gather}
    \biggl\langle \hat{p}_r^2\,\hat{\tau}_c\,\frac{1}{\hat{m}^2}\,\hat{\tau}_c\biggr\rangle_\Psi 
    \geq 
    \frac{2}{3m_r^2}\,\biggl\langle \hat{p}_r^2\,\hat{\tau}_c^2\biggr\rangle_\Psi.
\end{gather}
Finally, we assume positive correlations between $\hat{p}_r^2$ and
$\hat{\tau}_c^2$, so that  
\[ 
\Bigl\langle \hat{p}_r^2\,\hat{\tau}_c^2\Bigr\rangle_\Psi \ge
 \Bigl\langle \hat{p}_r^2\Bigr\rangle_\Psi \,\Bigl\langle\hat{\tau}_c^2\Bigr\rangle_\Psi .
\]
This is physically justified: the correlation between the clock and the rod
arises from a coupling between the c.o.m.\ momentum $\hat{p}_r$ and the
internal Hamiltonian $\hat{H}_c$ (through the mass–energy $\hat{m}$).
Uncertainty in the momentum induces decoherence in the internal-energy basis,
thereby reducing the precision of the clock (for an explicit example, see
Appendix~\ref{app:clock-precision}). 

Since we are not interested in kinematical drift effects, we consider the rest frame, where the average momentum vanishes, so that $\langle \hat{p}_r^2\rangle_\Psi = \Delta_\Psi^2\hat{p}_r$. Combining (i) with (ii) yields:
\begin{gather}
    \Delta_\Psi^2  \hat{\widetilde{x}}_{(x_0,\tau_0)}  \ge 
    \Delta_\Psi^2 \hat{x}_r + \Delta_\Psi^2 \hat{v}_r\,\tau_0^2 +
     \nonumber \\
      \frac{\Delta_\Psi^2 \hat{p}_r}{m_r^2}
       \Bigl(\frac{2}{3}\Delta_\Psi^2\hat{\tau}_c 
       - \frac{5}{4}\frac{\hbar^2}{m_r^2 c^4}\Bigr).
\end{gather}
We notice that in the low-energy regime, because
$\Delta_\Psi^2\hat{\tau}_c \gg \hbar^2/m_r^2c^4$, the negative term is
negligible. This leads to Eq.~\eqref{eq:spread-relational-position-rel-rod} in
the main text, namely
\begin{gather}\label{eq_appG:variance-last-inequality}
    \Delta_\Psi^2  \hat{\widetilde{x}}_{(x_0,\tau_0)}  \gtrsim
    \Delta_\Psi^2 \hat{x}_r + \Delta_\Psi^2 \hat{v}_r\,\tau_0^2 
     + \frac{2}{3}\frac{\Delta_\Psi^2 \hat{p}_r}{m_r^2}
      \Delta_\Psi^2\hat{\tau}_c .
\end{gather}
In the following we keep this term explicit until the very end, and show that this approximation is well justified. \\

\paragraph*{Minimum spread}

Using the commutation relations between $\hat{x}_r$, $\hat{v}_r$, and
$\hat{p}_r$, Eq. \eqref{eq_appG:variance-last-inequality} becomes
\begin{gather}
     \Delta_\Psi^2  \hat{\widetilde{x}}_{(x_0,\tau_0)}  \ge 
    \Delta_\Psi^2 \hat{x}_r 
     + \left(\frac{\hbar}{2}\,\langle \hat{m}^{-1}\rangle_\Psi\,\tau_0\right)^2
       \frac{1}{\Delta_\Psi^2 \hat{x}_r}\nonumber\\
     + \left(\frac{\hbar}{2m_r}\right)^2
       \!\left(\frac{2}{3}\Delta_\Psi^2\hat{\tau}_c- \frac{5}{4}\frac{\hbar^2}{m_r^2 c^4}\right)
       \frac{1}{\Delta_\Psi^2 \hat{x}_r}.
\end{gather}
Using $\langle \hat{m}^{-1}\rangle_\Psi =
\overline{m}_r^{-1}\big(1+\Delta_\Psi^2 \hat{H}_c/m_r^2 c^4\big)$ and the
time–energy uncertainty relation
$\Delta_\Psi\hat{\tau}_c\,\Delta_\Psi \hat{H}_c \ge \hbar/2$, we obtain
\begin{gather}
\Delta_\Psi^2  \hat{\widetilde{x}}_{(x_0,\tau_0)} \ge 
    \Delta_\Psi^2 \hat{x}_r 
     + \left(\frac{\hbar}{2\overline{m}_r}\tau_0\right)^2
       \Biggl[ 1 + \left(\frac{\hbar/m_r c^2}{2\Delta_\Psi\hat{\tau}_c}\right)^2\Biggr]
       \frac{1}{\Delta_\Psi^2 \hat{x}_r}
      \nonumber \\+\left(\frac{\hbar}{2m_r}\right)^2
       \!\left(\frac{2}{3}\Delta_\Psi^2\hat{\tau}_c- \frac{5}{4}\frac{\hbar^2}{m_r^2 c^4}\right)
       \frac{1}{\Delta_\Psi^2 \hat{x}_r}.
       \label{eq:spread-both-dX-dT}
\end{gather}
Minimizing Eq. \eqref{eq:spread-both-dX-dT} with respect to $\Delta_\Psi^2 \hat{x}_r$ gives
\begin{gather}
        \Delta_\Psi^2 \hat{\widetilde{x}}_{(x_0,\tau_0)} \ge 
        \frac{\hbar}{\overline{m}_r}|\tau_0| \cdot \nonumber\\
       \cdot \sqrt{1
        + 2\!\left(\frac{\hbar/2m_r c^2}{\Delta_\Psi\hat{\tau}_c}\right)^2
        + \frac{2}{3}\!\left(\frac{\Delta_\Psi\hat{\tau}_c}{|\tau_0|}\right)^2
        - \frac{5}{4}\!\left(\frac{\hbar/m_r c^2}{|\tau_0|}\right)^2}.
\end{gather}
The minimum with respect to $\Delta_\Psi\hat{\tau}_c$ is attained for
$(\Delta_\Psi\hat{\tau}_c)^2 \sim |\tau_0|\hbar/(m_r c^2)$, which is physically meaningful only when $|\tau_0|\gg\hbar/m_r c^2$. In this regime, the proper time uncertainty can be much larger than $\hbar/m_r c^2$ but still small compared to $|\tau_0|$. Explicitly, we find:
\begin{gather}
       \Delta_\Psi^2  \hat{\widetilde{x}}_{(x_0,\tau_0)}  \ge 
        \frac{\hbar}{\overline{m}_r}|\tau_0|
        \sqrt{1+\frac{2}{\sqrt{3}}\frac{\hbar/m_r c^2}{|\tau_0|}
        -\frac{5}{4}\left(\frac{\hbar/m_r c^2}{|\tau_0|}\right)^2}.
\end{gather}
Since $|\tau_0|\gg\hbar/m_r c^2$, the negative correction is subleasing. Expanding to first order gives Eq.\eqref{eq:rel-pos-spread-final} in the main text, namely 
\begin{gather}
       \Delta_\Psi^2  \hat{\widetilde{x}}_{(x_0,\tau_0)}  \gtrsim 
        \frac{\hbar}{\overline{m}_r}|\tau_0| 
        + \frac{1}{\sqrt{3}}\frac{\hbar^2}{m_r^2 c^2}.
\end{gather}
Reversing the order of minimizations yields the same result.

\end{document}